\documentclass[showkeys]{revtex4-2}

\usepackage{graphicx}
\usepackage{dcolumn}
\usepackage{bm}
\usepackage{hyperref}
\hypersetup{colorlinks=true, allcolors={blue}}
\usepackage{booktabs}

\newcommand{\bgso}{Ba$_2$GdSbO$_{6}$}
\newcommand{\sgso}{Sr$_2$GdSbO$_{6}$}

\begin{document}

\preprint{APS/123-QED}

\title{Frustrated Gd$^{3+}$ Double Perovskites as High-Performance Magnetocaloric Materials for Sub-100 mK Adiabatic Demagnetization Refrigeration}

\author{Tim Treu}
 \email{tim1.treu@uni-a.de}
\author{Marvin Klinger}%

\author{Christian Heil}

\author{Vinícius E. S. Frehse}
\affiliation{%
 Experimental Physics VI, Center for Electronic Correlations and Magnetism, Institute of Physics, University of Augsburg, 86159 Augsburg, Germany
}%

\author{Mamoun Hemmida}

\author{Hans-Albrecht Krug von Nidda}
\affiliation{Experimental Physics V, Center for Electronic Correlations and Magnetism, University of Augsburg, 86159 Augsburg, Germany}%

\author{Anton Jesche}
\affiliation{%
 Experimental Physics VI, Center for Electronic Correlations and Magnetism, Institute of Physics, University of Augsburg, 86159 Augsburg, Germany
}%

\author{Alexander A. Tsirlin}
\affiliation{Felix Bloch Institute for Solid State Physics, University of Leipzig, 04103 Leipzig, Germany}%

\author{Philipp Gegenwart}

 \email{philipp.gegenwart@physik.uni-augsburg.de}
\affiliation{%
 Experimental Physics VI, Center for Electronic Correlations and Magnetism, Institute of Physics, University of Augsburg, 86159 Augsburg, Germany
}%

\date{\today}

\begin{abstract} \noindent
Achieving temperatures below 100\,mK is essential for advancing quantum technologies and exploring fundamental quantum phenomena. While paramagnetic salts have traditionally enabled adiabatic demagnetization refrigeration (ADR), their limitations have driven the search for more effective alternatives. In this work, we present Gd$^{3+}$-based double perovskites, \bgso{} and \sgso{}, as high-performance magnetocaloric materials. Starting ADR from 2\,K and 5\,T, these compounds reach 67\,mK and 68\,mK in small finite magnetic fields, and 83\,mK and 78\,mK in zero field, respectively, which are the lowest reported ADR temperatures for Gd$^{3+}$ magnets (S~=~7/2) under these conditions. The frustrated geometry and the complex interplay of exchange and dipolar interactions suppress their antiferromagnetic ordering temperatures to 100 and 166\,mK for \bgso{} and 190\,mK for \sgso{}, while maintaining an outstanding entropy density of 189\,mJ\,K$^{-1}$\,cm$^{-3}$ and 201\,mJ\,K$^{-1}$\,cm$^{-3}$, respectively. Notably, \bgso{} sustains sub-100\,mK cooling even at finite fields of 0.5\,T, making it particularly promising for practical ADR applications.
\end{abstract}

\keywords{Adiabatic demagnetization refrigeration; Double-perovskite; Magnetocaloric effect; sub-Kelvin cooling; Magnetic frustration}

\maketitle

\section{Introduction} \noindent
Cooling represents a pivotal component within the domain of modern technology, encompassing diverse fields such as satellite sensors, medical imaging, and quantum computing. Adiabatic demagnetization refrigeration (ADR) is an effective way of achieving low temperatures below 1 K. Recent developments led a renewed interest in magnetocalorics~\cite{Moya2020}. Especially the temperature range of $^3$He-cooling (0.3 to 2\,K) has been identified as an ideal playground for ADR in order to replace this scarce and expensive fossil resource. Groundbreaking was the identification of frustrated magnets as promising candidates for cryogen-free magnetic refrigeration due to their suppressed ordering temperatures, highly degenerate ground states and high entropy densities \cite{Zhitomirsky2003}. In zero magnetic field, strongly frustrated magnets have a large, macroscopic degeneracy of their classical ground state arising from the condition that the frustrated unit is not sufficient to satisfy all microscopic degrees of freedom \cite{Ramirez1994}. As a result strongly frustrated magnets may remain disordered with finite entropies at temperatures well below their Curie-Weiss temperature, which gauges the energy scale of magnetic couplings~\cite{Zhitomirsky2003}. From the material's perspective, this allows a much denser packing of the magnetic ions and, therefore, a much higher magnetic entropy compared to ordinary paramagnetic salts that were previously used in ADR applications. Since then many frustrated but also non-frustrated magnets have been identified as possible materials for low-temperature ADR~\cite{Tokiwa2021, Delacotte2022, Xu2022, Yang2024, Manvell2025, Guchhait2025, Wang2024, Xu2024, Wang2025, Gong2026, Ashtar2026}. However, most of these materials fall short of reaching the sub-100\,mK temperatures required for many applications. \\
Apart from the well-known paramagnetic salts like CMN \cite{Fisher1973}, CPA \cite{Daniels1954} and FAA \cite{Vilches1966}, only few materials demonstrated temperatures below 100 mK in ADR. Prominent examples are the Yb-based compounds Yb$_3$Ga$_5$O$_{12}$ \cite{PaixaoBrasiliano2020}, KYb$_3$F$_{10}$ \cite{Xu2025}, KBaYb(BO$_3$)$_2$ \cite{Tokiwa2021} and (K,Na)YbP$_2$O$_7$ \cite{Arjun2023a}. All of these materials contain the pseudospin-1/2 Yb$^{3+}$ ions resulting in weak interactions and, therefore, very low magnetic ordering temperatures, which is one of the prerequisites for reduced ADR base temperatures. However, the maximum entropy, which is available for cooling, is limited to $R\ln(2)$. By contrast, Gd compounds have a three times higher magnetic entropy density due to the Gd$^{3+}$ $S$ = 7/2 spins, thus leading to enhanced magnetic volume density and potentially superior ADR performance. This can seen by comparing isostructural Yb and Gd compounds, such as KBa(Yb/Gd)(BO$_3$)$_2$ \cite{Tokiwa2021, Jesche2023} and Na(Yb/Gd)P$_2$O$_7$ \cite{Arjun2023a, Telang2025}. In these materials, the superior ADR performance of the Gd-compounds is limited only by their higher ordering temperatures, therefore limiting the base temperature of the refrigerant. Here, we avoid this limitation by introducing strongly frustrated three-dimensional (3D) networks built exclusively by Gd$^{3+}$ and showing low magnetic ordering temperatures. To this end, we use the face-centered cubic (fcc) lattice that can be seen as a network of edge-sharing tetrahedra, with the tetrahedron itself representing a three-dimensional frustrated configuration \cite{Ramirez1994}. Antiferromagnetic (AFM) nearest-neighbor (NN) interactions render fcc magnets strongly frustrated, although longer-range interactions \cite{Haar1962,Balla2020,Oitmaa2023} as well as thermal and quantum fluctuations \cite{Yildirim1998,Schick2020,Schick2022} usually stabilize magnetic order. \\
Experimentally, fcc spin lattice is realized in Gd-based double-perovskite compounds that have been proposed for ADR applications~\cite{Koskelo2022, Koskelo2023a, Koskelo2023}, but not yet tested in practical cooling setups. Their magnetic ordering temperatures have not been determined either. In this work, we experimentally demonstrate the outstanding ADR performance of \bgso{} and \sgso{}, driven by their high magnetic entropy density, weak AFM exchange couplings combined with competing dipolar couplings of similar strength and strongly suppressed magnetic ordering to temperatures of 100 and 166\,mK for \bgso{} and 190 mK for \sgso{}, accompanied by an exceptional entropy density of 189\,mJ\,K$^{-1}$\,cm$^{-3}$ and 201\,mJ\,K$^{-1}$\,cm$^{-3}$, respectively, surpassing many state-of-the-art mK magnetocaloric materials.
When ADR is initiated from 2\,K and 5\,T, these compounds achieve temperatures as low as 67\,mK and 68\,mK in small applied magnetic fields, and 83\,mK and 78\,mK in zero field, for \bgso{} and \sgso{} respectively. These temperatures represent the lowest reported ADR temperatures for Gd$^{3+}$ magnets under such conditions. These findings establish both compounds as high performance magnetocaloric materials, offering high entropy density and low ordering temperatures for next-generation ultra-low-temperature cooling systems.
 
\section{Results and Discussion}
\subsection{Crystal Structure}

\begin{figure*}[t]
\centerline{\includegraphics[width=0.95\linewidth]{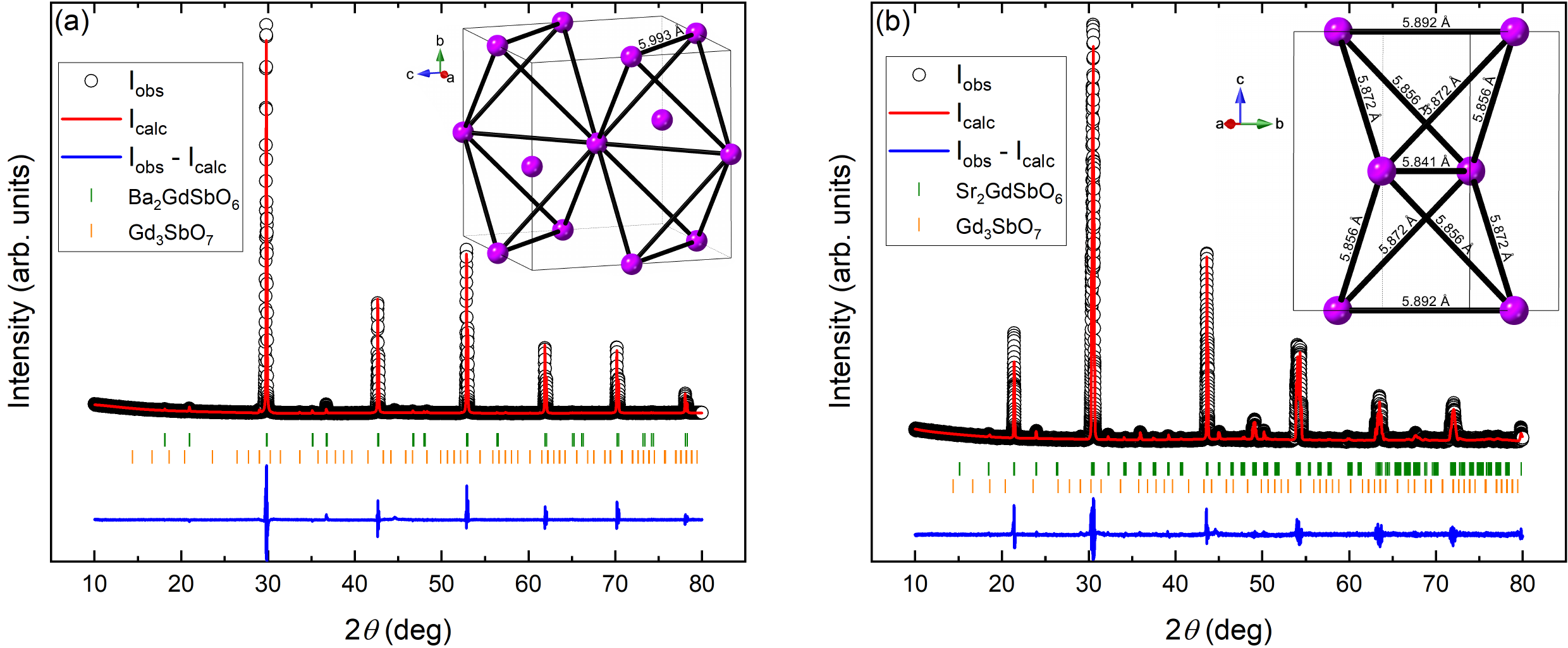}}
\caption{SPXRD pattern of Ba$_2$GdSbO$_6$ (a) and Sr$_2$GdSbO$_6$ (b) collected at 300\,K. The insets depict the Gd$^{3+}$-ion tetrahedral coordination within the respective unit cells, with bond distances given in \AA{}. \label{fig:xrd}}
\end{figure*}

\noindent Room temperature synchrotron powder X-ray diffraction (SPXRD) data confirm the formation of nearly phase-pure Ba$_2$GdSbO$_6$ and Sr$_2$GdSbO$_6$. The observed structures align with those previously reported in the literature \cite{Casado1984, Karunadasa2003, WongNg2014, Orayech2017, Koskelo2022}. Both compounds exhibit full rocksalt ordering of the magnetic Gd$^{3+}$ and nonmagnetic Sb$^{5+}$ ions on the B sites, resulting in alternating corner-sharing GdO$_6$ and SbO$_6$ octahedra in a double perovskite arrangement \cite{WongNg2014}. This ordering arises from the significant differences in charge and ionic radii between the cations \cite{Koskelo2022}. \\
Ba$_2$GdSbO$_6$ crystallizes in a cubic structure with space group $Fm\overline{3}m$, forming a frustrated network with uniform tetrahedra of Gd$^{3+}$ ions. The room-temperature SPXRD pattern, along with Rietveld refinement results and the Gd-ion arrangement in the unit cell, is shown in Figure~\ref{fig:xrd} (a). The diffraction data are well-described by the reported unit cell parameter~\cite{Koskelo2022}. However, minor additional peaks reveal trace amounts of Gd$_3$SbO$_7$ impurity (less than 1 wt.\,\%, as determined by two-phase refinement, referencing~\cite{Hinatsu2009} with a goodness of fit $\chi^2 = 3.62$), which shows AFM ordering at 2.6\,K~\cite{Hinatsu2009}. The refined lattice parameter $a = 8.4799$\,\AA{} and the NN Gd-Gd distance of 5.996\,\AA{} are consistent with literature values~\cite{Casado1984, Koskelo2022}. \\
In contrast, Sr$_2$GdSbO$_6$ adopts a monoclinic $P2_1/n$ structure, where the Gd$^{3+}$ tetrahedra deviate from regular geometry. A two-phase refinement of the measured pattern, using references from~\cite{Hinatsu2009, Koskelo2022}, reveals an impurity content of $<$0.5~wt.\% (see Figure~\ref{fig:xrd}(b)) with $\chi^2 = 4.70$. The refined lattice parameters ($a = 5.8501$\,\AA{}, $b = 5.8932$\,\AA{}, $c = 8.2997$\,\AA{}, and $\beta = 90.221^\circ$) agree well with published data~\cite{WongNg2014, Orayech2017, Koskelo2022}. The inset of Figure~\ref{fig:xrd} (b) illustrates the Gd-ion positions and the four distinct nearest-neighbor Gd-Gd distances, confirming the weakly distorted fcc geometry.

\subsection{Magnetic Properties}\label{sec:magnetic properties}

\begin{figure*}[htb]
\centerline{\includegraphics[width=0.85\linewidth]{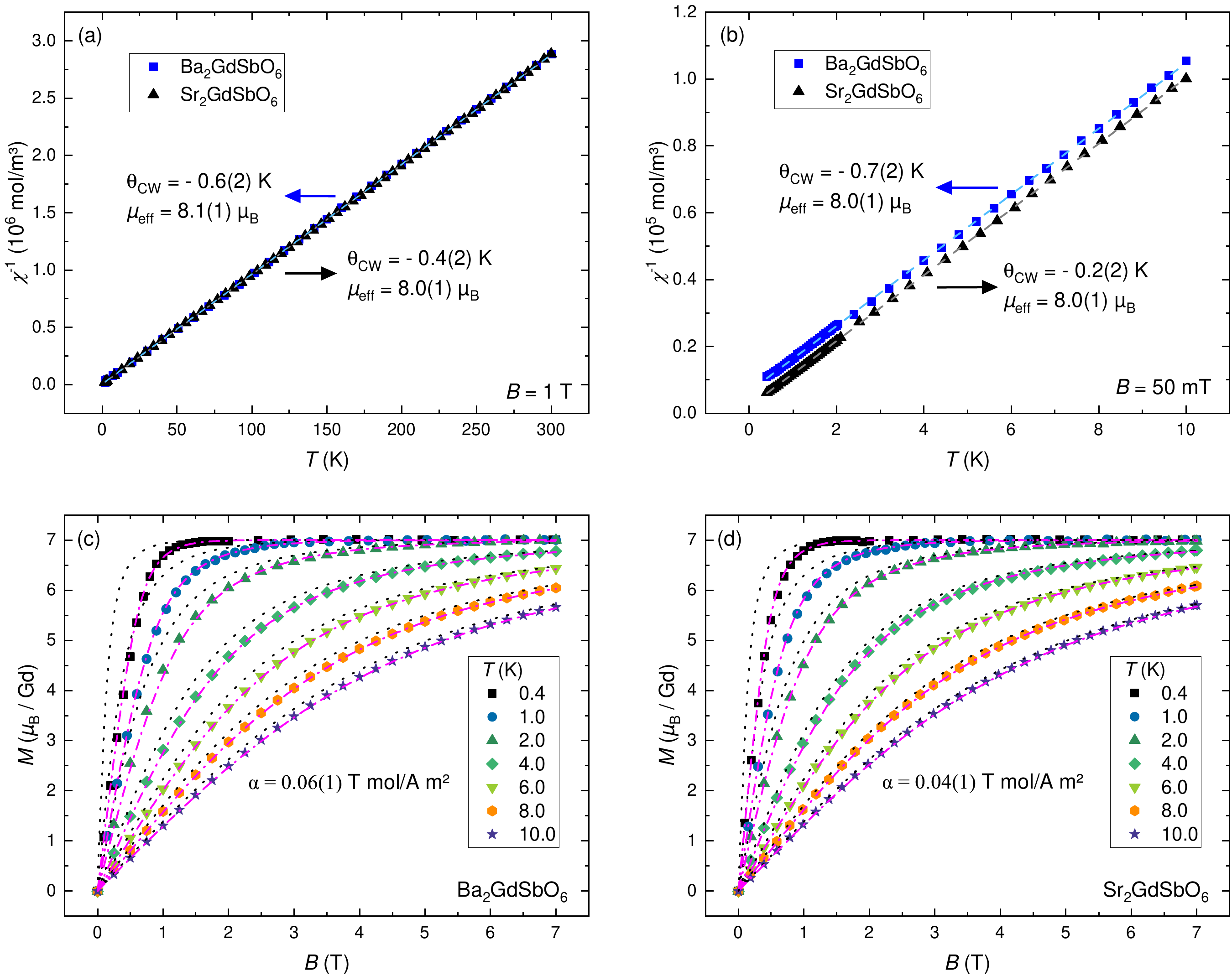}}
\caption{Magnetic susceptibility $\chi$ = $\mu_0 M/B$ of Ba$_2$GdSbO$_6$ and Sr$_2$GdSbO$_6$. Curie Weiss behaviour in 1/$\chi$ is observed over a wide temperature range in the high (a) and low temperature regime (b). To enhance visual clarity, only every fifth data point is plotted in the high temperature regime. Dashed lines are the result of fitting to the Curie-Weiss law (Eq. \ref{sup:eq:CW}). Isothermal magnetization of Ba$_2$GdSbO$_6$ (c) and Sr$_2$GdSbO$_6$ (d) expressed in $\mu_\mathrm{B}$ per Gd$^{3+}$. Every tenth data point is plotted for readability. The black dotted lines represent the calculated moments from Equation \ref{sup:eq:freeIon} for a free ion and the dashed pink lines depict a mean-field approximation as outlined in Equation \ref{sup:eq:meanField}.\label{fig:magnetic properties}}
\end{figure*}

\noindent The temperature dependence of the magnetic susceptibility, $\chi$~=~$\mu_0 M/B$, for Ba$_2$GdSbO$_6$ and Sr$_2$GdSbO$_6$ is shown in Figure \ref{fig:magnetic properties} (a) and (b). In the high-temperature region $T$~=~10~-~300\,K, both materials show Curie-Weiss behaviour with an effective magnetic moment of $\mu_{\mathrm{eff}}$ = 8.1(1)\,$\mu_\mathrm{B}$ for Ba$_2$GdSbO$_6$ and $\mu_{\mathrm{eff}}$~=~8.0(1)\,$\mu_\mathrm{B}$ for Sr$_2$GdSbO$_6$, in good agreement with the value of the free Gd$^{3+}$ ion ($\mu_{\mathrm{eff}}~=~g_\mathrm{J}\sqrt{J(J+1)}\mu_\mathrm{B}~=~7.94\,\mu_\mathrm{B}$). Negative Curie-Weiss temperatures $\Theta_{\mathrm{CW}}$ of -0.6(2)\,K (Ba$_2$GdSbO$_6$) and -0.4(2)\,K (Sr$_2$GdSbO$_6$) indicate AFM interactions in good agreement with \cite{Koskelo2022}. \\
The low-temperature behavior of $\chi(T)$ reveals a linear evolution of 1/$\chi$ persisting down to 0.4\,K, with no deviation observed at the lowest measurable temperature. The Curie Weiss fit in the region $T$~=~0.4~-~10\,K returns an effective moment of $\mu_{\mathrm{eff}}$ = 8.0(1)\,$\mu_\mathrm{B}$ for both compounds and a Curie-Weiss temperature of $\Theta_{\mathrm{CW}}$~=~-0.7(2)\,K for Ba$_2$GdSbO$_6$ and -0.2(2)\,K for Sr$_2$GdSbO$_6$. \\
The mean-field expression for the Curie-Weiss temperature,
\begin{equation}\label{eq:cw}
 \Theta_{\rm CW}=-\frac{zS(S+1)}{3}\sum_i J_i,
\end{equation}
returns $\Theta_{\rm CW}=-63\bar J$ for the fcc lattice where $\bar J$ is the average nearest-neighbor coupling and z = 12 the number of nearest neighbors. Using the values from the high-temperature Curie-Weiss fit, we determine $\bar J=9.5$\,mK in \bgso{} and 6.3\,mK in \sgso{}, suggesting that the symmetry lowering reduces the nearest-neighbor coupling by about 30\,\%. \\
The isothermal magnetization $M(B)$ is depicted in Figure \ref{fig:magnetic properties} (c) and (d) for Ba$_2$GdSbO$_6$ and Sr$_2$GdSbO$_6$ respectively. For the largest applied fields, the measured moments at temperatures of 2\,K and below approach their saturation value and are in a good agreement with the saturation moment of the free Gd$^{3+}$ ion $\mu_{\mathrm{sat}}~=~7\,\mu_{\mathrm{B}}$. It is evident that the free-ion model, as expressed in Equation \ref{sup:eq:freeIon}, is not able to adequately describe the observed curves. This deviation is attributable to weak but non-negligible interactions between the magnetic ions, which results in a lower increase of the magnetization in increasing applied fields in both compounds. \\
A better fit is obtained using the mean-field expression for interacting magnetic ions, Eq.~\ref{sup:eq:meanField}, that returns consistent values of the interaction constant $\alpha$ for each of the two compounds. The lower magnitude of $\alpha$ in \sgso{} compared to \bgso{} confirms the 30\% reduction in the exchange couplings upon replacing Ba with Sr. The values of $\alpha$ and $\Theta_{\rm CW}$ are related by $\alpha=\mu_0\Theta_{\rm CW}/C$ where $C$ is the Curie constant. Using $C=7.88$\,cm$^3$\,K/mol for Gd$^{3+}$ and Eq.~\ref{eq:cw}, we find $\bar J=6.0$\,mK in \bgso{} and 4.0\,mK \sgso{}, in a acceptable agreement with the values obtained from the Curie-Weiss fits.

\subsection{Specific Heat}

\begin{figure*}[htb]
\centerline{\includegraphics[width=0.85\linewidth]{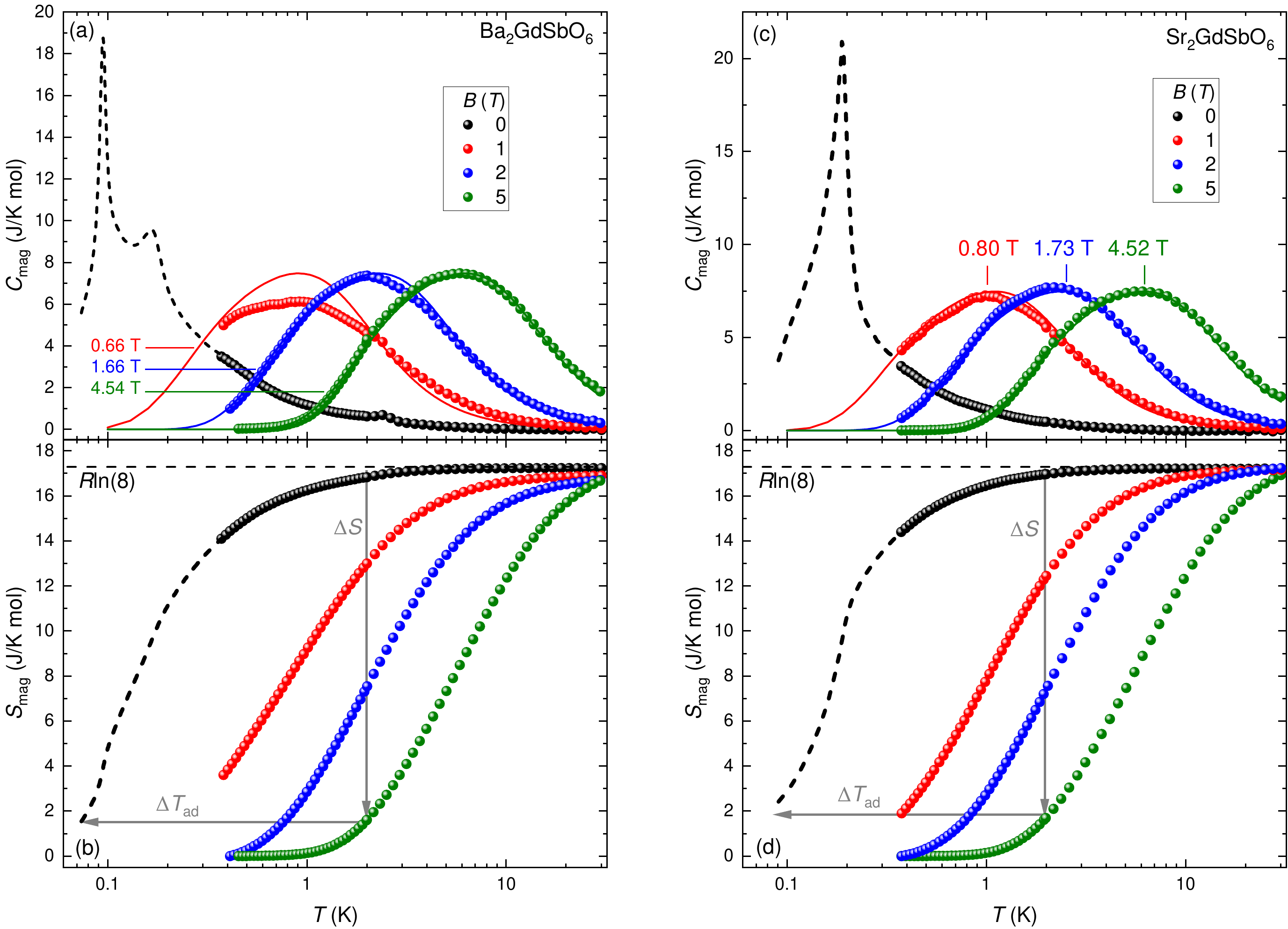}}
\caption{Temperature-dependent magnetic specific heat $C_{\mathrm{mag}}(T)$ (upper panels) and entropy $S_{\mathrm{mag}}(T)$ (lower panels) of (a) \bgso{} and (b) \sgso{}. The dashed lines indicate data calculated from the ADR warming curves (see Section \ref{sec:ADR}). The colored lines illustrate model calculations based on Equation \ref{eq:Schottky}, utilising the specified effective field values. Grey arrows indicate the potential to cool to below 100\,mK during an ADR process. \label{fig:HC_Entropy}}
\end{figure*}

\noindent The temperature-dependent magnetic contribution to the specific heat, $C_{\mathrm{mag}}(T)$, for Ba$_2$GdSbO$_6$ and Sr$_2$GdSbO$_6$ is shown in Figure \ref{fig:HC_Entropy} (a) and (b). A small anomaly in the zero-field specific heat curve of Ba$_2$GdSbO$_6$ is attributed to AFM ordering of the Gd$_3$SbO$_7$ impurity phase at 2.6\,K \cite{Hinatsu2009}. For low temperatures in zero field below 0.4\,K, the microcalorimeter measurements were extended by means of ADR warmup analysis, revealing peaks that will be discussed in detail in Section \ref{sec:ADR}. Additional measurements of $C_{\mathrm{mag}}(T)$, were conducted under applied magnetic fields of $B$ = 1, 2, and 5\,T. At these elevated fields, the magnetic ordering is completely suppressed, and $C_{\mathrm{mag}}(T)$ instead exhibits Schottky-like anomalies, with the peak shifting toward higher temperatures with increasing magnetic field. These anomalies were analyzed using Equation \ref{eq:Schottky}, with the internal fields and magnetic exchange interactions accounted for by introducing an effective field $B_{\mathrm{eff}}$ acting on each magnetic ion. With the exception of the 1~T curve for \bgso{}, where the effect of magnetic interactions may be still non-negligible, the model describes the data quite well. The $B_{\mathrm{eff}}$ values consistently fell below the applied field, indicating the presence of weak but non-negligible AFM exchange interactions. \\
The magnetic entropy curves $S_{\mathrm{mag}}(T)$ calculated from $C_{\mathrm{mag}}(T)$ following Eq. \ref{sup:eq:Entropy} are plotted in the lower panels of Figure \ref{fig:HC_Entropy}. Since a significant amount of entropy for $B$~=~0 and 1\,T is not accessible from $C_{\mathrm{mag}}(T)$, we utilized the entropy differences derived from isothermal magnetization analysis (see Section \ref{sup:sec:HC}). There is good agreement to the expected saturation value $R$ln(8)~=~17.3\,J/K\,mol. Gray arrows mark the entropy reduction and respective expected ADR performance starting at 2\,K. 

\subsection{Quasi-Adiabatic Demagnetization Refrigeration}\label{sec:ADR}

\begin{figure*}[htb]
\centerline{\includegraphics[width=0.9\linewidth]{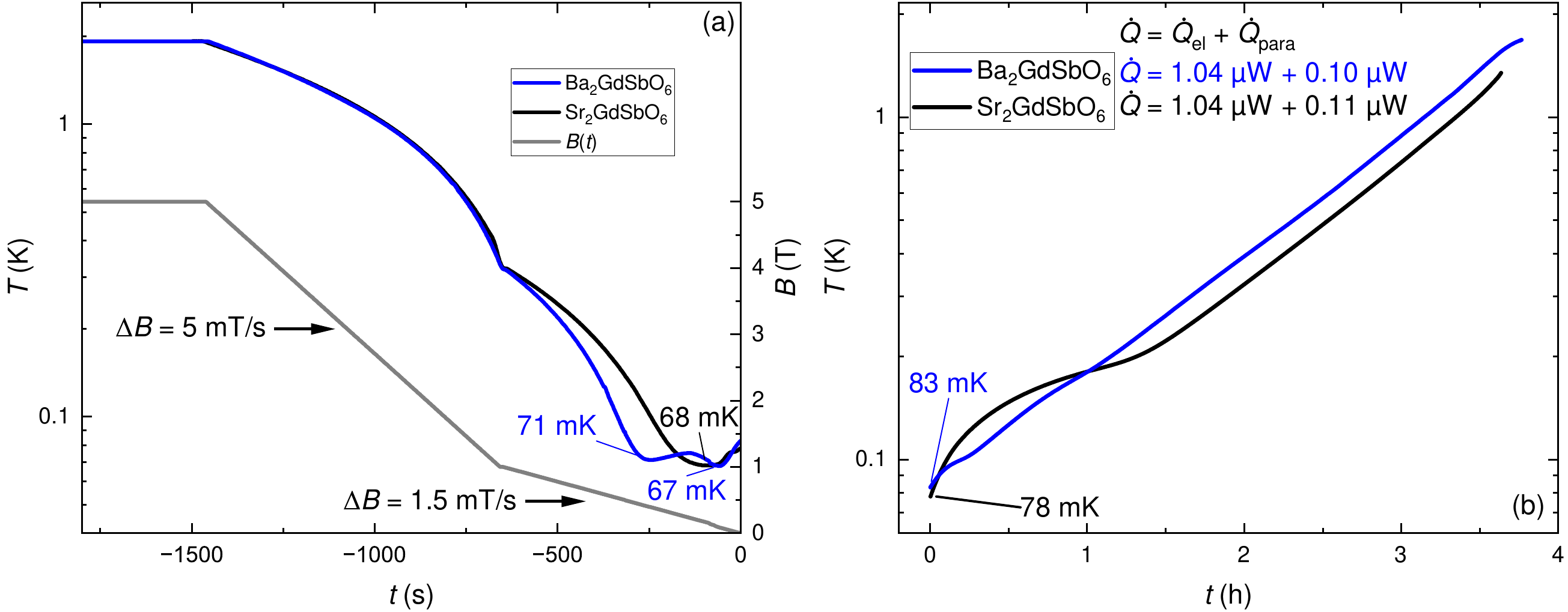}}
\caption{(a) ADR cooling curves starting from $T$ = 2 K and $B$ = 5 T. The ramp rate was reduced below 1 T to improve temperature measurement accuracy. \bgso{} exhibits two minima at 71 mK (0.4 T) and 67 mK (0.1 T), while \sgso{} shows a single anomaly at 68 mK and 0.15 T. (b) Warming curves for both compounds after reaching zero field at $t$ = 0, with phase transitions visible as kinks in the curves. Note that the short warmup time is caused by the deliberate introduction of significant electrical heating $Q_\mathrm{el}$ (see Section \ref{sup:sec:ADR})\label{fig:ADR_curves}}
\end{figure*}

\begin{figure*}[htb]
\centerline{\includegraphics[width=0.61\linewidth]{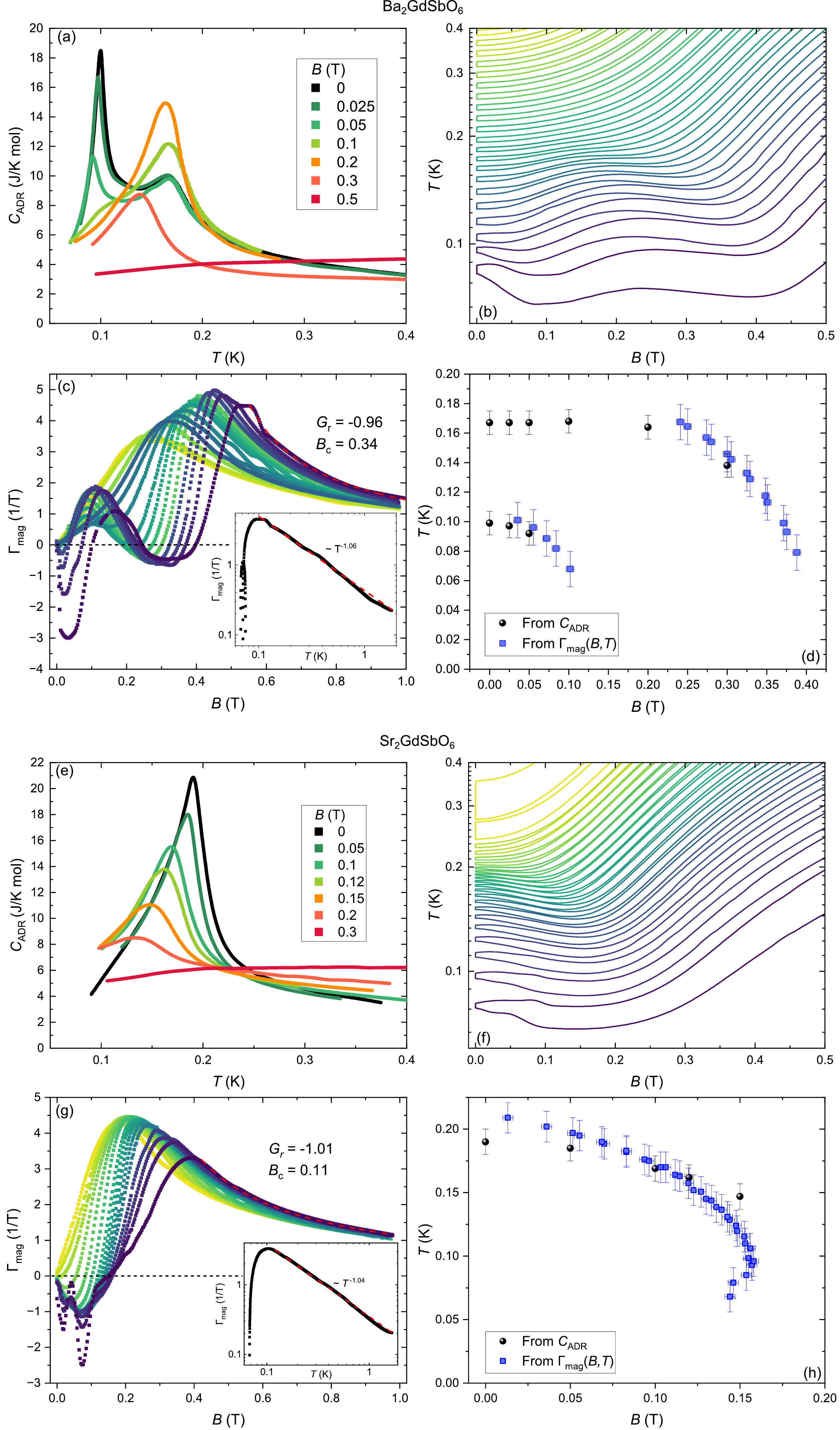}}
\caption{Heat capacity $C_\mathrm{ADR}$ of \bgso{} (a) and \sgso{} (e) derived from ADR warming curves using Eq. \ref{sup:eq:ADR}. (b), (f) $T(B)$ plots of \bgso{} (a) and \sgso{} recorded by sweeping the applied field between 0 and 1 T, starting from 2\,K and 5\,T. The minima in the traces indicate magnetic phase boundaries. (c), (g) Magnetic Gr\"uneisen parameter $\Gamma_\mathrm{mag}$, calculated from the $T(B)$ traces using Eq. \ref{eq:Grüneisenparameter}. The inset shows the temperature dependence of $\Gamma_\mathrm{mag}$. (d), (h) Tentative phase diagram, where the peak positions from (a), (e) and the zero crossings from (c), (g) are marked. \label{fig:ADR}}
\end{figure*}

\noindent The ADR performance of both compounds was evaluated in a home-built setup (see Section \ref{sup:sec:ADR}). The magnetic field was ramped from the initial conditions of  $T~=~2$\,K and $B~=~5$\,T,T to zero or finite values, with the temperature $T(t)$ recorded during cooling (while ramping the field) and during warmup (after turning on the heater). The results are shown in Figure \ref{fig:ADR_curves}. While the cooling behavior is very similar down to 1\,T, at lower temperatures deviations occur. \sgso{} reaches a minium at 68\,mK at roughly 0.15\,T and reaches 78\,mK in zero field. \bgso{} undergoes two minima reaching 71\,mK and 67\,mK at 0.4\,T and 0.1\,T respectively and reaches 83\,mK in zero field. These are the lowest base temperatures for Gd$^{3+}$ magnets reported to date, even surpassing many Yb$^{3+}$-compounds (see Section \ref{sec:comparison}) under these conditions. DFT calculations (see Section \ref{sup:sec:DFT}) reveal that the symmetry lowering in the double-perovskite structure reduces nearest-neighbor exchange couplings in \sgso{} compared to \bgso{} in agreement with our thermodynamic analysis in Section \ref{sec:magnetic properties}. Remarkably, though, both compounds show very similar minimum cooling temperatures despite the 30\,\% difference in $\bar J$, likely due to the presence of dipolar couplings and weak crystal electric field effects in \sgso{} as evidenced by ESR measurements (see Section \ref{sup:sec:ESR}). \\
The warmup curves show two intersections, due to anomalies indicative of magnetic ordering. As the onset of magnetic order restricts the base temperature of ADR, the type of order and the magnitude of the ordering temperature are critical microscopic parameters for ADR materials. Being a witness of magnetic ordering transition, the specific-heat anomaly does not discriminate between antiferromagnetic, ferromagnetic, or more complex types of order. To shed more light on this issue, we derived the heat capacity $C_\mathrm{ADR}$, following established methods (see Section \ref{sup:sec:ADR}), with results shown in Figure \ref{fig:ADR} (a) for \bgso{} and (e) for \sgso{}. For \bgso{} in zero field, a sharp peak is observed at 100(5)\,mK, with a broader, smaller peak at 166(10)\,mK. In small finite fields, the sharp peak shifts to lower temperatures and is gradually suppressed by 0.2\,T. The broader peak remains largely unchanged up to 0.05\,T but increases above 0.1\,T, with no clear temperature shift. At 0.3\,T, the peak is drastically reduced, shifted to lower temperatures, and fully suppressed. Such a field evolution is typical of AFM transitions. For \sgso{} in zero field, a single sharp peak is observed at 190(7)\,mK. It shifts to lower temperatures with applied field, again suggesting AFM nature of the transition. The transition is fully suppressed at 0.3\,T and a Schottky anomaly appears. \\
Additionally, we investigated thermodynamic signatures of the phase transitions via approximately isentropic ($dS$ = 0) temperature traces $T(B)$, by repeatedly ramping the magnetic field between 0 and 1\,T (see Figs. \ref{fig:ADR} (b) and (f)). The minimum in $T(B)$ indicate positions of entropy accumulation which are generically expected at the boundary of a second-order magnetic phase transition near a quantum critical point (QCP) \cite{Zhu2003,Garst2005, Zhou2026}. Following the positions of these minima in the $T$-$B$ phase space allows to draw the boundary of the AFM phases. Further information is obtained from the magnetic Gr\"uneisen parameter $\Gamma_\mathrm{mag}$~\cite{Zhu2003}, which we calculate from the isentropes using 
\begin{equation} \label{eq:Grüneisenparameter}
\Gamma_\mathrm{mag} = \frac{1}{T}\left(\frac{dT}{dB}\right)_S.
\end{equation}
The field dependence of $\Gamma_\mathrm{mag}$, derived from downward magnetic field sweeps is shown in Figure \ref{fig:ADR} (c) and (g). The color of the various curves is similar to the respective temperature traces in pannels (b) and (e). A sign change from negative to positive $\Gamma_\mathrm{mag}(B)$ indicates a second-order phase transition, signified by a local entropy maximum~\cite{Garst2005}. In the case of \bgso{}, two zero crossings are clearly visible, shifting to lower fields for higher-lying isentropes. In contrast, \sgso{} exhibits only one zero crossing. Notably, at low fields, a local maximum without zero crossing is observed, which quickly smears out in higher lying isentropes, suggesting the possible presence of a second magnetically ordered phase at very low temperatures, below the limit of these measurements. \\
For a field-tuned QCP, a divergence of $\Gamma_\mathrm{mag}$ from large fields in the approach of the critical field is expected~\cite{Garst2005}. This divergence was fitted assuming the critical behavior $\Gamma_\mathrm{mag}$~=~-$G_\mathrm{r}$/$\mid$$B$-$B_\mathrm{c}$$\mid$, yielding the value of $G_\mathrm{r} \approx -1$ for both compounds and the critical fields of 0.34\,T and 0.11\,T for \bgso{} and \sgso{}, respectively. Additionally, the fit of the temperature dependence of $\Gamma_\mathrm{mag}$ (see inset of Figs. \ref{fig:ADR} (c) and (g)) in the vicinity of the critical field using $\Gamma \propto 1/T^{1/\nu z}$ returns $\nu z=1$ for both compounds, where $\nu$ is the critical exponent of the correlation length, and $z$ is the dynamical critical exponent. According to Ref.~\cite{Zhu2003}, the pre-factor $G_r$ is related to the critical exponents as follows,
\begin{equation}
 G_r=\nu\left(\frac{d-y_0z}{y_0}\right)
\end{equation}
where $\nu$ and $z$ have been defined above, whereas $d=3$ is the spatial dimension, and $y_0$ is the critical exponent of the specific heat. Given $\nu z=1$, the condition $G_r=-1$ can be satisfied with $y_0\rightarrow\infty$ only. This is the known behavior of a dilute Bose gas of magnons~\cite{Garst2005} that are gapped in the fully polarized state above the critical field and condense upon crossing the quantum critical point and entering the AFM ordered state. The thermodynamic properties of both \bgso{} and \sgso{} are thus fully consistent with the scenario of AFM order that occurs at low temperatures and low fields and gets suppressed as the field increases. 
Combining the positions of zero crossings in $\Gamma_\mathrm{mag}(T,B)$ with the peak positions from $C_{\mathrm{ADR}}(T)$, field-temperature phase diagrams were constructed. In the case of \bgso{}, two distinct phase boundaries are visible. In \sgso{}, only one clear phase boundary was observed. It shows non-monotonic behavior at temperatures below 0.1\,K, resulting from the shift of the zero crossing in $\Gamma_\mathrm{mag}$ and is consistent with the obtained critical field. The lower critical field of \sgso{} is in agreement with the weaker exchange couplings in this compound compared to \bgso{}. \\
Turning to the origin of magnetic order, we first note that two types of magnetic interactions, exchange and dipolar, could occur in Gd$^{3+}$ double perovskites. A purely dipolar fcc magnet develops ferromagnetic order~\cite{Luttinger1946} with the Curie temperature of about 90\,mK~\cite{Bouchaud1993} for the Gd-based double perovskites considered in this work. By contrast, AFM exchange couplings result in magnetic frustration that impedes magnetic order, although at very low temperatures the order eventually sets in as a result of interactions beyond nearest neighbors and/or quantum effects. The behavior of \bgso{} and \sgso{} is overall consistent with the latter scenario, namely, magnetic order appears at temperatures well below $\Theta_{\rm CW}$ indicating strong magnetic frustration. This order is AFM in nature, as confirmed by the presence of the magnetic transition in finite fields and by the critical scaling of the magnetic Gr\"uneisen parameter. Both compounds show non-negligible AFM exchange couplings according to our analysis of the magnetic susceptibility and magnetization data as well as DFT. On the other hand, the dipolar couplings are comparable in energy to the exchange couplings. Therefore, the Gd$^{3+}$ double perovskites combine two mechanisms of frustration, the geometrical frustration of the fcc lattice and the competition between exchange and dipolar couplings that stabilize different types of order in the fcc network. \\
Another interesting observation is the occurrence of two consecutive magnetic transitions in \bgso{}. Most of the fcc antiferromagnets show one magnetic transition only. The appearance of a second transition can be a result of symmetry lowering, as in Sr$_2$YRuO$_6$ where partial magnetic order appearing at the upper transition precedes complete magnetic order that sets in below the second transition~\cite{Granado2013}. Two consecutive transitions may also occur in an undistorted fcc antiferromagnet exemplified by Ba$_2$YOsO$_6$~\cite{Kermarrec2015}, but the nature of the second ordered state and the origin of the two transitions have not been resolved in the literature. \bgso{} adds another material to this conundrum and raises the interesting question of the interplay of exchange and dipolar couplings in fcc magnets.

\subsection{Exceptional Sub-Kelvin Cooling Performance} \label{sec:comparison}

\begin{table*}[t]
  \centering
  \caption{Comparison of key properties of mK ADR materials, sorted by volumetric entropy density $S_\mathrm{GS}$/vol., based on published molar volumes. Where applicable, $T_\mathrm{m}$ refers to the magnetic transition temperature (or short-range order, marked with *), $T_\mathrm{min}$ to the lowest reported ADR temperature,  $S_\mathrm{GS}$ to the ground state multiplet entropy, and $R$ to the universal gas constant. The abbreviations stand for: CMN~=~Mg$_3$Ce$_2$(NO$_3$)$_{12}\cdot$24H$_2$O (cerium magnesium nitrate), FAA~=~NH$_4$Fe(SO$_4$)$\cdot$12H$_2$O (ferric ammonium alum), CPA~=~KCr(SO$_4$)$\cdot$12H$_2$O (chromium potassium alum) and MAS~=~Mn(NH4)$_2$(SO$_4$)$_2\cdot$6H$_2$O (manganese ammonium sulphate).} \label{tab:ADR comparison}
  \begin{tabular*}{\textwidth}{@{\extracolsep\fill}lccccc@{\extracolsep\fill}}
    \toprule
    Compound & $S_\mathrm{GS}$ & $S_\mathrm{GS}$/vol. & $T_\mathrm{m}$ & ADR $T_\mathrm{min}$ & Ref. \\
     & & (mJ K$^{-1}$ cm$^{-3}$) & (mK) & (mK) & \\
    \midrule
    CMN & $R$ln(2) & 16 & 2 & & \cite{Fisher1973}\\
	CPA & $R$ln(4) & 42 & 10 & & \cite{Daniels1954}\\
	FAA & $R$ln(6) & 53 & 30 & & \cite{Daniels1954}\\
    NaYbP$_2$O$_7$ & $R$ln(2) & 64 &  & 45 & \cite{Arjun2023a}\\
	KBaYb(BO$_3$)$_2$ & $R$ln(2) & 64 & 9 & 40 & \cite{Tokiwa2021,Jesche2023}\\
    Ba$_3$GdB$_9$O$_{18}$ & $R$ln(8) & 75 & 368 & 94 & \cite{Klinger2025}\\
	MAS & $R$ln(6) & 70 & 170 & & \cite{Vilches1966} \\
    Ba$_3$GdBiPbB$_4$O$_{13}$ & $R$ln(8) & 86 & 190* & 69 & \cite{Ashtar2026}\\
    Ba$_3$GdB$_3$O$_9$ & $R$ln(8) & 122 & 435 & 119 & \cite{Klinger2025}\\
	Yb$_3$Ga$_5$O$_{12}$ & $R$ln(2) & 124 & 180 & 100 & \cite{PaixaoBrasiliano2020} \\
	YbNi$_{1.6}$Sn & $R$ln(2) & 148 & 140* & 116 & \cite{Gruner2024}\\
    Yb$_2$Be$_2$GeO$_7$ & $R$ln(2) & 149 & 380* & 95 & \cite{Liu2024}\\
    KYb$_3$F$_{10}$ & $R$ln(2) & 156 & $<$ 50 & 27 & \cite{Xu2025}\\
    LiYb$_{0.9}$Gd$_{0.1}$F$_4$ & 0.1$R$ln(8)+0.9$R$ln(2) & 163 & 85 & 160 & \cite{Xu2025b} \\
	YbBO$_3$ & $R$ln(2) & 181 & 399 & 202 & \cite{Sala2023, Treu2025}\\
	\textbf{Ba$_2$GdSbO$_6$} & $R$ln(8) & 189 & 95 & 67 & [this work]\\
	KBaGd(BO$_3$)$_2$ & $R$ln(8) & 192 & 263 & 122 & \cite{Jesche2023}\\
	\textbf{Sr$_2$GdSbO$_6$} & $R$ln(8) & 201 & 190 & 68 & [this work]\\
	NaGdP$_2$O$_7$ & $R$ln(8) & 210 & 570 & 215 & \cite{Telang2025}\\
    Eu$_2$ZnGe$_2$OS$_6$ & $R$ln(8) & 212 & 280 & 144 & \cite{Gong2026}\\
	Gd$_{0.1}$Yb$_{0.9}$F$_3$ & 0.1$R$ln(8)+0.9$R$ln(2) & 244 & 180 & & \cite{Xu2024} \\
	Gd$_3$Ga$_5$O$_{12}$ & $R$ln(8) & 363 & 900* & 350 & \cite{Daudin1982,Kleinhans2023}\\
    GdCO$_3$F & $R$ln(8) & 432 & 950 & & \cite{Xu2024a}\\
    Gd$_{0.7}$Yb$_{0.3}$F$_3$ & 0.7$R$ln(8)+0.3$R$ln(2) & 466 & 700 & &  \cite{Xu2024} \\
    Gd(OH)F$_2$ & $R$ln(8) & 498 & 500 & & \cite{Xu2022} \\
    Gd$_3$BWO$_9$ & $R$ln(8) & 502 & 1080 & 151 & \cite{Wang2025, Song2025} \\
	Gd$_{9.33}$[SiO$_4$]$_6$O$_2$ & $R$ln(8) & 509 & 500 & 300 & \cite{Yang2024}\\
    \bottomrule
  \end{tabular*}
\end{table*}

\begin{figure*}[ht]
\centerline{\includegraphics[width=0.8\linewidth]{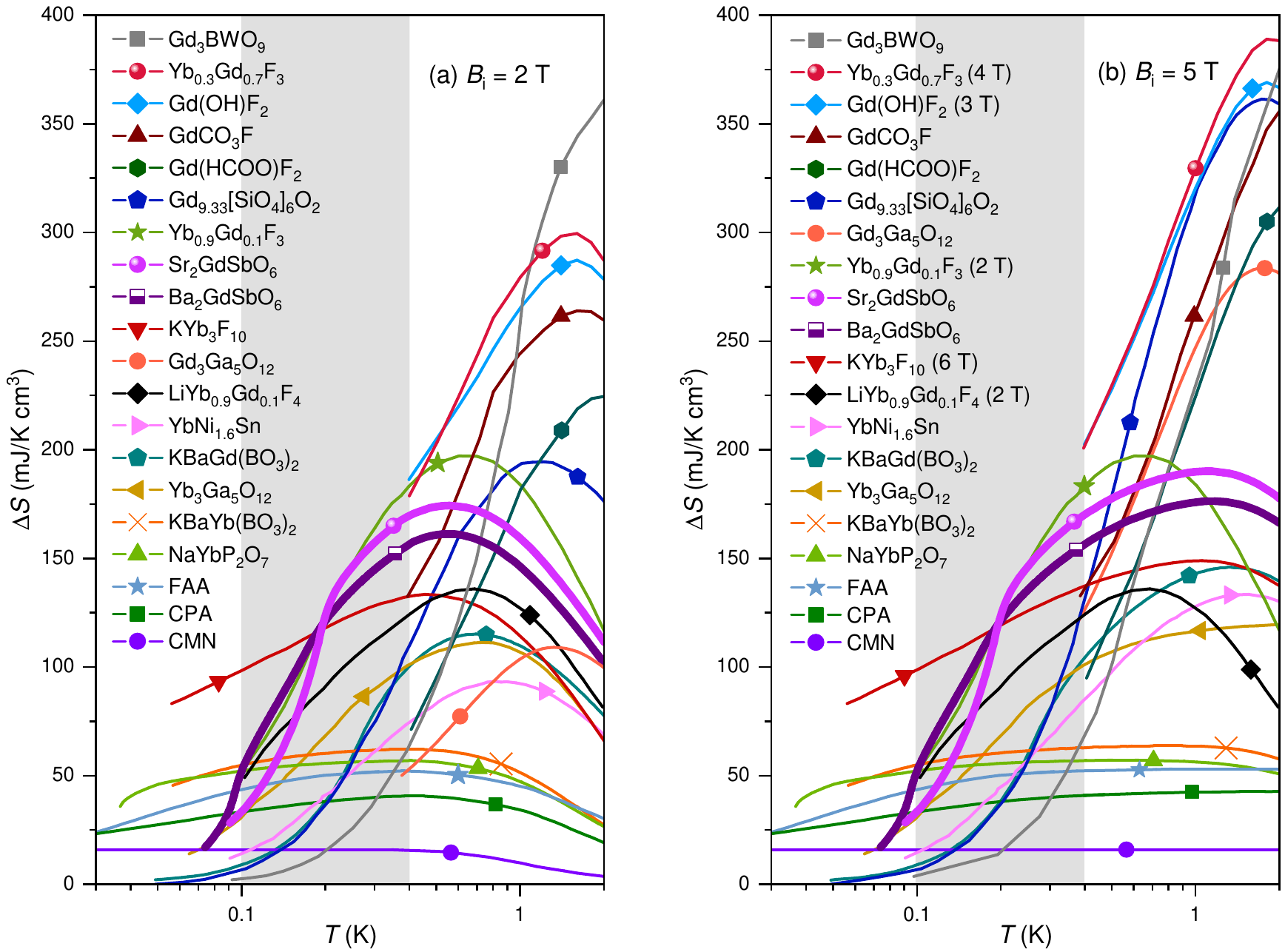}}
\caption{Volumetric entropy density increment, $\Delta S(T)~=~S(T, 0)~-~S(T, B_\mathrm{i})$, as a function of temperature for mK ADR materials at initial magnetic fields of (a) $B_\mathrm{i}$ =  2 T and (b) $B_\mathrm{i}$ = 5 T (where available). The curves of the compounds discussed in this work are bolded for clarity and the region of interest between 0.1 and 0.4\,K is shaded in grey. Calculations are based on specific heat or entropy data from the literature (referenced in Table \ref{tab:ADR comparison}), with interpolation or extrapolation used as needed.} \label{fig:ADR_Comparison}
\end{figure*}

\noindent When compared to traditional and state-of-the-art mK magnetocaloric materials (see Table \ref{tab:ADR comparison} and Figure \ref{fig:ADR_Comparison}), both \bgso{} and \sgso{} exhibit a superior cooling effect at low temperatures. Among the few published Gd$^{3+}$-based compounds that reach temperatures below 100\,mK, Ba$_3$GdBiPbB$_4$O$_{13}$ achieves 69\,mK starting from 2\,K and 9\,T \cite{Ashtar2026}, while Ba$_3$GdB$_9$O$_{18}$ reaches 94\,mK starting from 2\,K and 5\,T \cite{Klinger2025}. In these compounds, the large separation between Gd$^{3+}$ ions minimizes magnetic interactions, whereas magnetic frustration additionally impedes magnetic order. Despite their more than twofold reduced entropy density, these materials do not achieve lower temperatures than \bgso{} and \sgso{}. Replacing Gd$^{3+}$ with Yb$^{3+}$ should generally reduce the ADR base temperature, but the Gd$^{3+}$ double perovskites presented in this work in fact outperform many of the reported Yb$^{3+}$ materials, including YbBO$_3$ \cite{Treu2025}, Yb$_2$Be$_2$GeO$_7$ \cite{Liu2024}, YbNi$_{1.6}$Sn \cite{Gruner2024} and Yb$_3$Ga$_5$O$_{12}$ \cite{PaixaoBrasiliano2020} while having a increased cooling power. Although base temperature can be influenced by thermal insulation and other aspects of the experimental setup, the thermodynamic characteristics, including the magnetic ordering temperature, also show a clear advantage of the Gd$^{3+}$ double perovskites over many of the Yb-based materials. \\
For paramagnetic salts and other non-frustrated systems, the magnetic entropy density generally correlates positively with cooling power in ADR, but this often comes at the cost of a higher base temperature \cite{Wikus2014}. However, as shown in Table \ref{tab:ADR comparison}, frustrated systems defy this trend as their ordering temperatures are significantly reduced. To explore this further, we calculated the volumetric entropy density increment, defined as $\Delta S(T)~=~S(T, 0)~-~S(T, B_\mathrm{i})$, for high-performing mK ADR materials at initial magnetic fields of $B_\mathrm{i}$ = 2 and 5\,T below 2\,K (see Figure \ref{fig:ADR_Comparison}). While fluoride-based materials exhibit the highest entropy density increments across most temperature ranges due to the extremely dense packing of magnetic ions on frustrated lattices, the double-perovskites investigated here outperform traditional compounds within the range of 0.1 to 0.4\,K. The current benchmark in this range is Yb$_3$Ga$_5$O$_{12}$, whose frustrated hyperkagome structure enables dense magnetic ion packing while suppressing long-range AFM ordering down to 54 mK \cite{Filippi1980, Raymond2024}. Our results demonstrate that both double-perovskite borates significantly exceed the cooling power of Yb$_3$Ga$_5$O$_{12}$, nearly matching the performance of leading low-temperature materials like Yb$_{0.9}$Gd$_{0.1}$F$_3$, while KYb$_3$F$_{10}$ shows the largest values below 190 mK. However, fluorides require complex synthesis involving highly toxic hydrofluoric acid  \cite{Xu2022, Xu2024a, Xu2025, Xu2025a, Xu2025b} while Gd double perovskites are non-toxic, can be easily synthesized and processed to pellets that provide comparable excellent mK ADR performance.

\section{Conclusion} \noindent
In this study, we synthesized polycrystalline \bgso{} and \sgso{} and investigated their magnetic and thermodynamic properties. We demonstrated that these Gd-based double perovskites achieve the lowest published ADR temperatures for Gd$^{3+}$ magnets, establishing them as high-performance ADR materials in the 100 to 400\,mK range. Their crystal structure ensures relatively large separations between Gd$^{3+}$ ions while maintaining outstanding entropy densities of 189\,mJ\,K$^{-1}$\,cm$^{-3}$ and 201\,mJ\,K$^{-1}$\,cm$^{-3}$, respectively. This unique performance arises from two concurrent mechanisms of frustration: the geometrical frustration of the fcc lattice and the competition between exchange and dipolar couplings of similar strength, which suppress magnetic order. Our thermodynamic measurements and ADR experiments uncover magnetic ordering transitions at 100\,mK and 166\,mK for \bgso{} and 190\,mK for \sgso{}, yet they still exhibit exceptional cooling performance below these ordering temperatures. Notably, \bgso{} achieves temperatures below 100\,mK even at finite fields of 0.5\,T, making it particularly suitable for applications in which mK temperatures shall be reached not only in zero-field but also in the presence of finite fields (for instance for investigating superconductors and their critical fields).

\section*{Funding} \noindent
The work was supported by the Deutsche Forschungsgemeinschaft (DFG, German Research Foundation), Grants No. 514162746 (GE 1640/11–1) and No. TRR 360–492547816.

\section*{Conflicts of Interest} \noindent
The authors declare no conflicts of interest.

\section*{Data Availability Statement} \noindent
The data that support the findings of this study are openly available at the following DOI: 10.5281/zenodo.22145694 

\bibliography{bibliography}

\appendix
\counterwithin{figure}{section}
\counterwithin{table}{section}
\renewcommand{\thefigure}{\thesection\arabic{figure}}
\renewcommand{\thetable}{\thesection\arabic{table}}
\section{Sample Preparation} \noindent
Polycrystalline Ba$_2$GdSbO$_6$ and Sr$_2$GdSbO$_6$ were synthesized by a conventional solid-state reaction of Gd$_2$O$_3$ (99.9\,\% ChemPur), Sb$_2$O$_3$ (99.99+\,\% ChemPur) and BaCO$_3$ (99.95\,\% Thermo Scientific), respectively, SrCO$_3$ (99.999\,\% ChemPur) \cite{Casado1984, Karunadasa2003, WongNg2014, Orayech2017, Koskelo2022}. The educts were meticulously ground in an agate mortar, transferred into an aluminium oxide crucible, and heated in a furnace in air three times at 1400\,°C for 24 hours, with intermediate grinding steps. Prior to weighing, all educts were preheated to remove moisture and carbonates. The same method was used to synthesize the doped polycrystalline compounds for ESR measurements, with the specific amount of Y$_2$O$_3$ being added. (99.9\,\% Alfa Aesar).

\section{Powder X-ray Diffraction} \noindent
The phase purity of the samples was confirmed using high-resolution synchrotron powder X-ray diffraction (SPXRD) measurements at the ID22 beamline of the ESRF (Grenoble, France). The experiments were conducted at room temperature with a wavelength of 0.40004\,\AA{}. Fine polycrystalline powder was loaded into a spinning quartz capillary. Rietveld refinements were performed using JANA2006 \cite{Jana} and Fullprof \cite{RodriguezCarvajal1993} and crystal structures were visualized with CrystalMaker\textsuperscript{\textregistered} \cite{CrystalMaker}.

\section{Calorimetric Measurements} \label{sup:sec:HC} 
\begin{figure}[htb]
\includegraphics[width=0.8\linewidth]{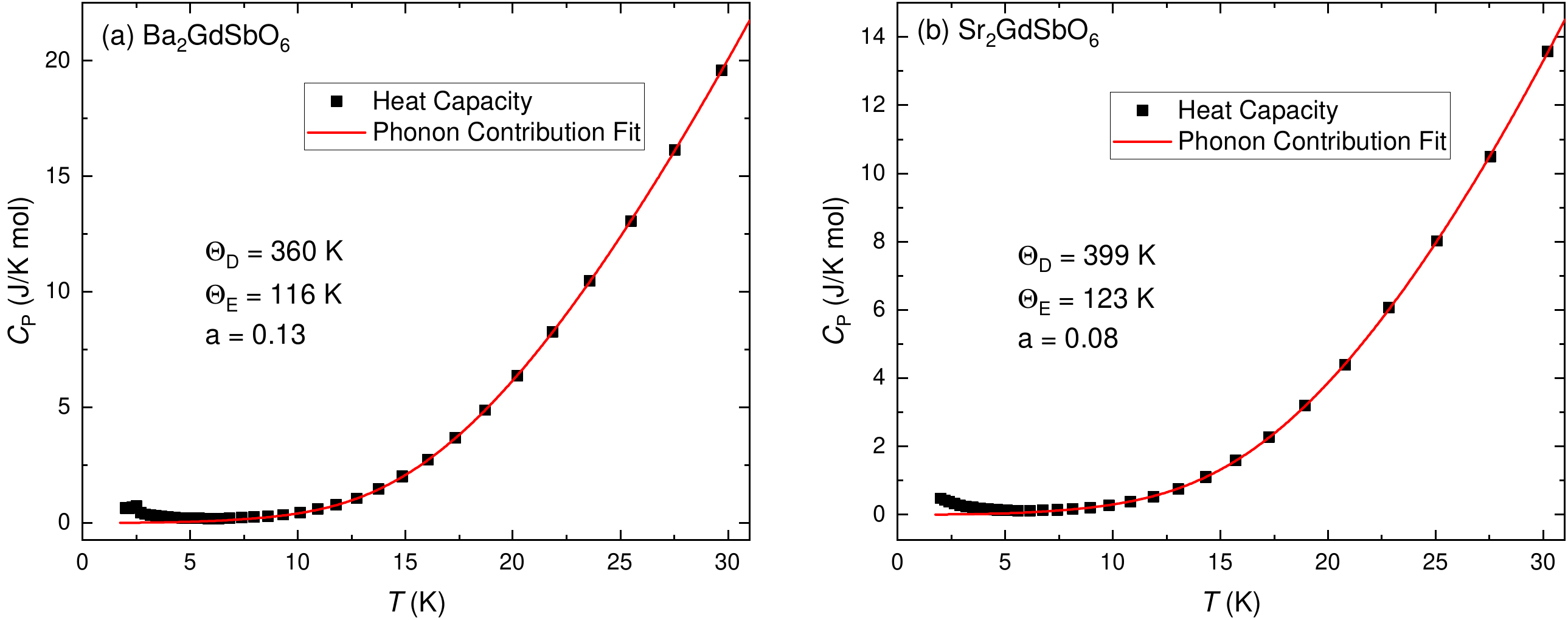}
\caption{Temperature dependence of $C_\mathrm{p}(T)$ in zero field for (a) \bgso{} and (b) \sgso{}. $C_\mathrm{lat}(T)$ was obtained by fitting $C_\mathrm{p}(T)$ with a Eq: \ref{sup:eq:HeatCapacityFit}. \label{sup:fig:Phonon_Contribution}}
\end{figure}

\begin{figure}[htb]
\includegraphics[width=0.8\linewidth]{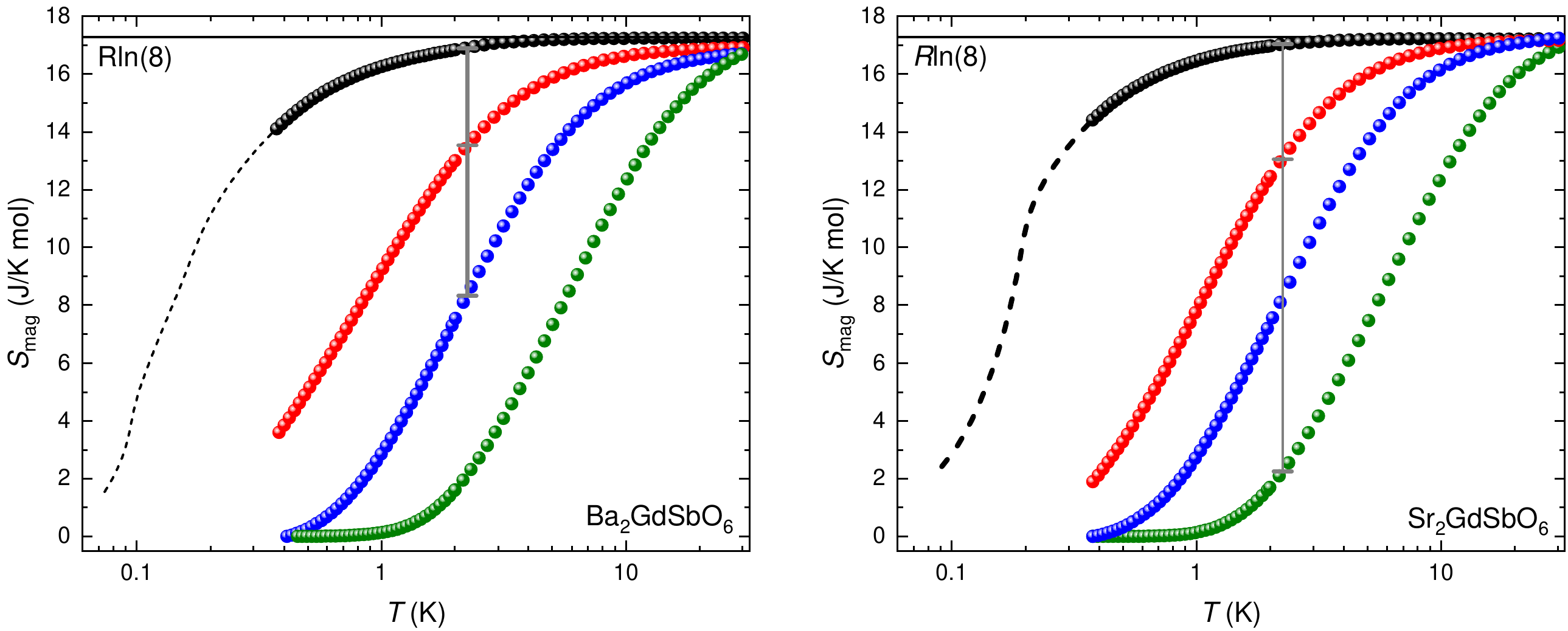}
\caption{Temperature and field dependence of the magnetic entropy $S_{\mathrm{mag}}(T)$. Grey vertical lines indicate the shift of the entropy curves for $B$~=~0 and 1\,T. \label{sup:fig:Entropy}}
\end{figure}

\noindent
The temperature-dependent specific heat, $C(T)$, was measured in the range of 0.4~-~30\,K in both zero and non-zero magnetic fields using the heat capacity option of a Dynacool Physical Property Measurement System (PPMS) manufactured by Quantum Design and equipped with a $^3$He option. In order to ensure strong thermal coupling, the sample powders were mixed with fine silver powder and pressed into 3\,mm pellets. For measurements employing the $^3$He option at temperatures below 2\,K, a small piece was cut from the pellet of each compound to account for the reduced cooling power. The samples were attached and thermally coupled to the stage with Apiezon N-grease. Addenda measurements of the sample platform and grease were calibrated at each temperature prior to measurement. The contribution of the silver powder was subtracted from the measured heat capacity $C_{\mathrm{tot}}$ using the data based on measurements of pure silver pellets of comparable size, to obtain the sample heat capacity $C_{\mathrm{p}}$. Finally, the magnetic specific heat $C_{\mathrm{mag}}$ was derived by subtracting the lattice contribution $C_{\mathrm{lat}}$ which was determined by using least-squares fits of the zero field $C_{\mathrm{p}}$ in the temperature range 13~-~30\,K to a combined Debye and Einstein model:
\begin{eqnarray} \label{sup:eq:HeatCapacityFit}
C_{\mathrm{lat}}(T) =&& (1-a) * \frac{9nRT^3}{\Theta_\mathrm{D}^3} \int_{0}^{\Theta_\mathrm{D}/T} \frac{x^4 e^x}{(e^x - 1)^2} \mathrm{d}x + a * 3nk_\mathrm{B}\left(\frac{\Theta_\mathrm{E}}{T}\right)^2\frac{e^{\frac{\Theta_\mathrm{E}}{T}}}{(e^{\frac{\Theta_\mathrm{E}}{T}}-1)^2},
\end{eqnarray}
where $R$ is the molar gas constant, $n$ is the number of atoms per formula unit, $k_\mathrm{B}$ is the Boltzmann constant, $\Theta_\mathrm{D}$ is the Debye temperature, $\Theta_\mathrm{E}$ is the Einstein temperature and $a$ is a scaling factor between the two models (see Figure \ref{sup:fig:Phonon_Contribution}). The obtained values of $\theta_\mathrm{D}$ are consistent with values reported in \cite{Koskelo2022}. \\ 
$C_{\mathrm{mag}}$ curves measured with applied external magnetic field were fitted with the general equation for the Schottky anomaly for a system of non-interacting magnetic dipoles (\cite{{Pobell2007}} Equation (9.15b)):
\begin{equation} \label{eq:Schottky}
 C_\mathrm{mag} = \frac{x^2R}{4}\sinh^{-2}\left( \frac{x}{2} \right) - \frac{x^2}{4}(2S+1)^2 \sinh^{-2}\left( \frac{x}{2}(2S+1) \right) \text{with} \quad x = \frac{\mu_\mathrm{B}gB_\mathrm{eff}}{k_\mathrm{B}T}.
\end{equation}
$B_{\mathrm{eff}}$ describes the contribution of internal fields and AFM exchange, $S$ stands for spin quantum number (since $L~=~0$ and so $J = S$) and $g$ for the Landé~factor. \\
From $C_{\mathrm{mag}}(T)$ the magnetic entropy is calculated by
\begin{equation}\label{sup:eq:Entropy}
    S_\mathrm{mag} =  S_0 + \int_{0.4\,\mathrm{K}}^{30\,\mathrm{K}} \frac{C_\mathrm{mag}}{T} \mathrm{d}T.
\end{equation}
$S_0$ represents a field-dependent constant, which is needed as not all of $C_{\mathrm{mag}}$ is recovered by the lowest measured temperature. Utilizing the Maxwell relation d$M$/d$T$~=~d$S$/d$B$, we obtain these offset constants by field-integration of the temperature derivative of the isothermal magnetization (see below). For \bgso{}, the entropy curves were adjusted based on the value of $\Delta S_{\mathrm{mag}}$ at 2.25\,K. The extracted values are: $\Delta S_{\mathrm{mag}}$~=~$S_{\mathrm{mag}}$(0\,T)~-~$S_{\mathrm{mag}}$(2\,T) =~8.9\,J/K\,mol and $\Delta S_{\mathrm{mag}}$~=~$S_{\mathrm{mag}}$(0\,T)~-~$S_{\mathrm{mag}}$(2\,T)~=~5.2\,J/K\,mol. The agreement of the saturation value of $S_{\mathrm{mag}}$(0\,T)~=~17.2\,J/K\,mol and $S_{\mathrm{mag}}$(0\,T) =~16.9\,J/K\,mol at $T$~=~30\,K to the expected value $R$ln(8)~= 17.3\,J/K\,mol is good. A similar approach was applied to \sgso{}, with entropy curves aligned based on $\Delta S_{\mathrm{mag}}$ at 2.25\,K. The values obtained are $\Delta S_{\mathrm{mag}}$ =~$S_{\mathrm{mag}}$(0\,T)~-~$S_{\mathrm{mag}}$(5\,T)~=~14.4\,J/K\,mol and $\Delta S_{\mathrm{mag}}$~=~$S_{\mathrm{mag}}$(0\,T)~-~$S_{\mathrm{mag}}$(5\,T) =~10.9\,J/K\,mol. Again, excellent agreement of the saturation of $S_{\mathrm{mag}}$(0\,T) =~17.2\,J/K\,mol and $S_{\mathrm{mag}}$(0\,T)~=~17.1\,J/K\,mol with the expected entropy was found. These shifts are marked by grey vertical lines in Figure \ref{sup:fig:Entropy}. Similar consistency was observed for $\Delta S_{\mathrm{mag}}$ curves at higher temperatures (3.25\,K and 4.25\,K), while for $T <$ 2\,K deviations become more pronounced due to increased uncertainties in d$M$/d$T$ arising from the approximation of the differential magnetization. \\
For the zero field curve, the specific heat obtained from the ADR warming curve (see section \ref{sec:ADR}) was used to extend the specific heat and entropy to below 0.4\,K.

\section{Magnetic measurements} \noindent
Magnetization measurements were conducted as a function of temperature ($T$~=~0.4~-~300\,K) and magnetic field (up to $B$~=~7\,T), employing an MPMS3 magnetometer from QuantumDesign, equipped with a $^3$He option. These measurements were performed on the same 3\,mm pellets as for the heat capacity measurements. For the measurements, the contributions from background signals, radial and axial offsets, sample geometry, demagnetization effects, mass uncertainty and magnetic field accuracy were carefully accounted for. \\
The temperature dependence of the magnetic susceptibility was fitted with the Curie-Weiss law
\begin{equation} \label{sup:eq:CW}
\chi(T) = C/(T-\Theta_\mathrm{CW}),
\end{equation}
from which the Curie-Weiss temperature $\Theta_\mathrm{CW}$ and the effective magnetic moment $\mu_{\mathrm{eff}} = \sqrt{\frac{3 k_\mathrm{B}C}{N_\mathrm{A}\mu_0}}$ was extracted. \\
The isothermal magnetization was fitted with a free ion model consisting of the Brillouin function
\begin{equation}\label{sup:eq:freeIon}
M(B) = gS \left[\frac{2S + 1}{2S}\coth\left(\frac{2S + 1}{2S} \frac{g \mu_\mathrm{B}  S  B}{k_\mathrm B T} \right)\right.
 \left.- \frac{1}{2S}\coth\left(\frac{g \mu_\mathrm{B} B}{2k_\mathrm B  T}\right)\right],
\end{equation}
where $S$ stands for spin quantum number and $g$ for the Landé~factor. Fixed values for $T$, $S = 7/2$ and $g = 2$ were used for the fitting. In addition, effect of magnetic couplings was included on the mean-field level using
\begin{equation} \label{sup:eq:meanField}
    M = \mu_\mathrm{sat}B_J\left[ \frac{g\mu_\mathrm{B}J}{k_\mathrm{B}T}\left(B+\alpha M \right) \right],
\end{equation}
where $\alpha$ denotes the mean field coupling constant. The obtained results are 0.06(1) and 0.04(1) T\,mol/A\,m$^2$ for \bgso{} and \sgso{} respectively (see Figure\ref{fig:magnetic properties}). From the $M(B)$ curves, experimental d$M$/d$T$ curves were obtained by 
\begin{equation} \label{sup:eq:dMdT}
\left(\frac{\partial M(T_0,B)}{\partial T}\right)_B \approx \frac{M(T_{i+1}, B) - M(T_i,B)}{T_{i+1} - T_i}.
\end{equation} 
This enables the calculation of the difference in magnetic entropy by utilising Maxwell relations, which result in
\begin{equation} \label{sup:eq:DeltaS}
\Delta S_m(T_0,B_{\mathrm{max}}) = \int_{0}^{B_{\mathrm{max}}} \left(\frac{\partial M(T_0,B)}{\partial T}\right)_B \mathrm{d}B.
\end{equation}
\begin{figure}[htb]
\includegraphics[width=0.85\columnwidth]{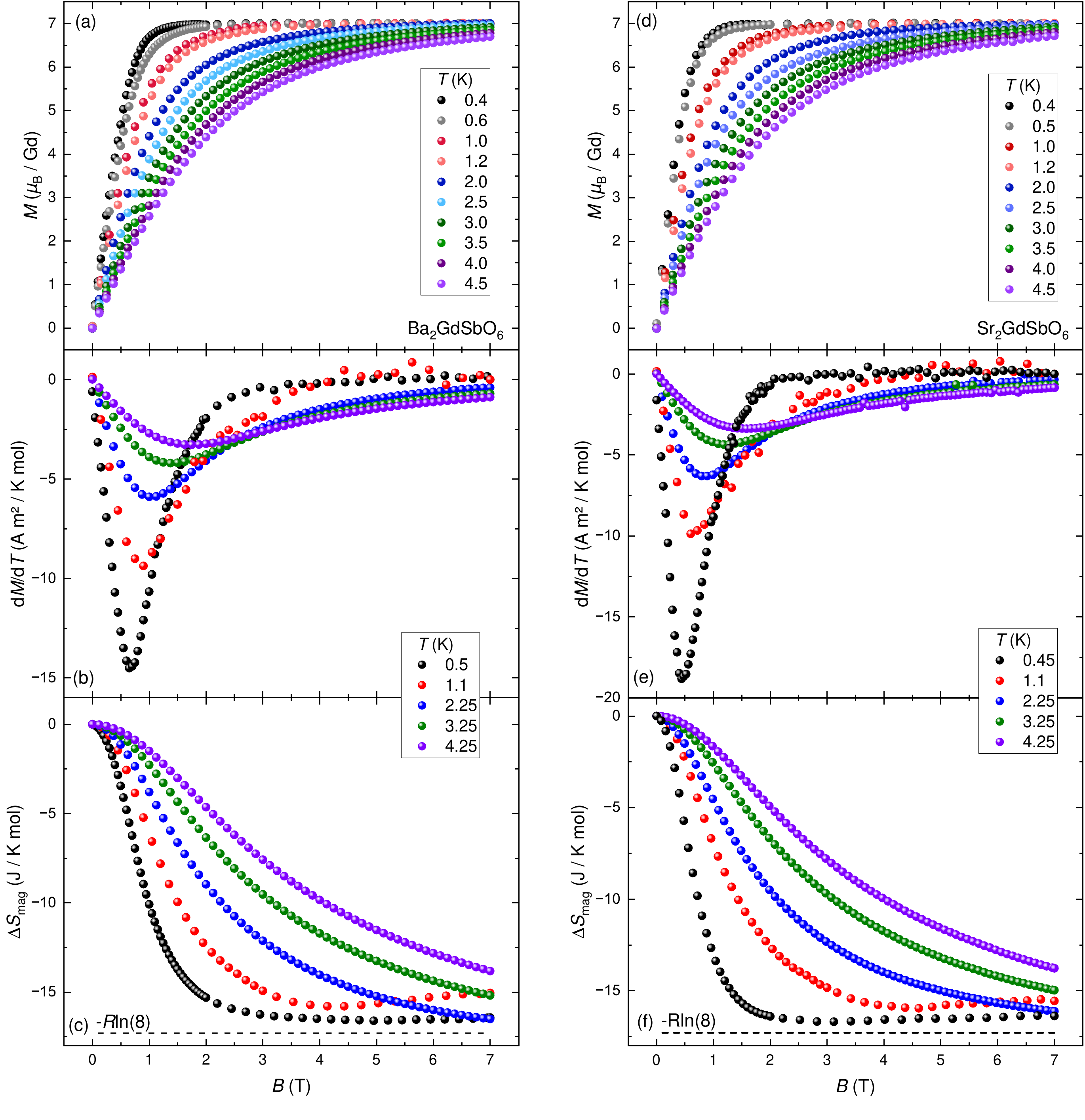}
\caption{Isothermal magnetization of Ba$_2$GdSbO$_6$ (a) and Sr$_2$GdSbO$_6$ (d) expressed in $\mu_\mathrm{B}$ per Gd$^{3+}$. (b)(e) Dependence of the magnetization’s temperature derivative with applied magnetic field, derived from Equation \ref{sup:eq:dMdT}, for Ba$_2$GdSbO$_6$ and Sr$_2$GdSbO$_6$ respectively. (c)(f) Magnetic entropy change as a function of applied field, obtained by Equation \ref{sup:eq:DeltaS}. In all panels every fifth data point is plotted for clarity. \label{sup:fig:Magnetic Properties}}
\end{figure} \\
Figure \ref{sup:fig:Magnetic Properties} (a) and (d) show the field dependence of several measured isothermal magentization curves. Figure \ref{sup:fig:Magnetic Properties} (b) and (e) show the change of $M$ with respect to temperature as a function of field. Adjacent $M(B)$ curves have been used to form pairs: 0.4\,K and 0.5\,K $\rightarrow$ d$M$/d$T$ at T~=~0.45\,K, 1.0\,K and 1.2\,K $\rightarrow$ d$M$/d$T$ at T~=~1.1\,K, 2.0\,K and 2.5\,K $\rightarrow$ d$M$/d$T$ at T~=~2.25\,K and similarly for d$M$/d$T$ at T~=~3.25\,K and 4.25\,K. For Ba$_2$GdSbO$_6$ the $M(B)$ curve at 0.6\,K was used instead of the one at 0.5\,K due to smaller noise. The two compounds display nearly identical behavior, with the maximum -d$M$/d$T$ observed at low temperatures due to faster saturation and diminished magnetization at higher temperatures. \\
For $\Delta S_{\mathrm{mag}}$ shown in Figure \ref{sup:fig:Magnetic Properties} (c) and (f), similar behaviour between the compounds is once again observed as the magnetic entropy decreases with increasing applied magnetic field. The most significant changes occur at the lowest temperatures, with $\Delta S_{\mathrm{mag}}~=~-16.6$\,J/K\,mol at $T~=~0.5$\,K for Ba$_2$GdSbO$_6$ and $\Delta S_{\mathrm{mag}}~=~-16.3$\,J/K\,mol at $T~=~0.45$\,K for Sr$_2$GdSbO$_6$. The slight increase in $\Delta S_{\mathrm{mag}}$ below 2\,K at higher fields and the resulting nonphysical crossing of curves is an artifact resulting from measurement uncertainties in $M(B)$. At elevated temperatures, $\Delta S_{\mathrm{mag}}$ continues to increase with the applied field, even at the highest fields studied, and gradually approaches the theoretical maximum value of $R \ln 8~=~17.3$\,J/K\,mol.

\section{Electron Spin Resonance Measurements} \label{sup:sec:ESR} \noindent
Electron spin resonance (ESR) measurements were conducted in a continuous wave mode spectrometer Bruker ELEXSYS E500 at X-band ($\nu~\approx$~9.35\,GHz) frequency over a temperature range of 4\,K to 300\,K, employing a continuous helium gas-flow cryostat (Oxford Instruments). ESR detects the microwave power $P$ absorbed by the sample from the transverse magnetic field as a function of the static magnetic field $B$, arising from magnetic dipole transitions between the Zeeman levels of electron spins. To enhance the signal-to-noise ratio, the derivative d$P$/d$B$ was recorded using lock-in detection with field modulation. The measurements were performed on pure polycrystalline powder samples, which were immobilized in a quartz tube using paraffin.
\begin{figure}[htb]
\includegraphics[width=0.8\linewidth]{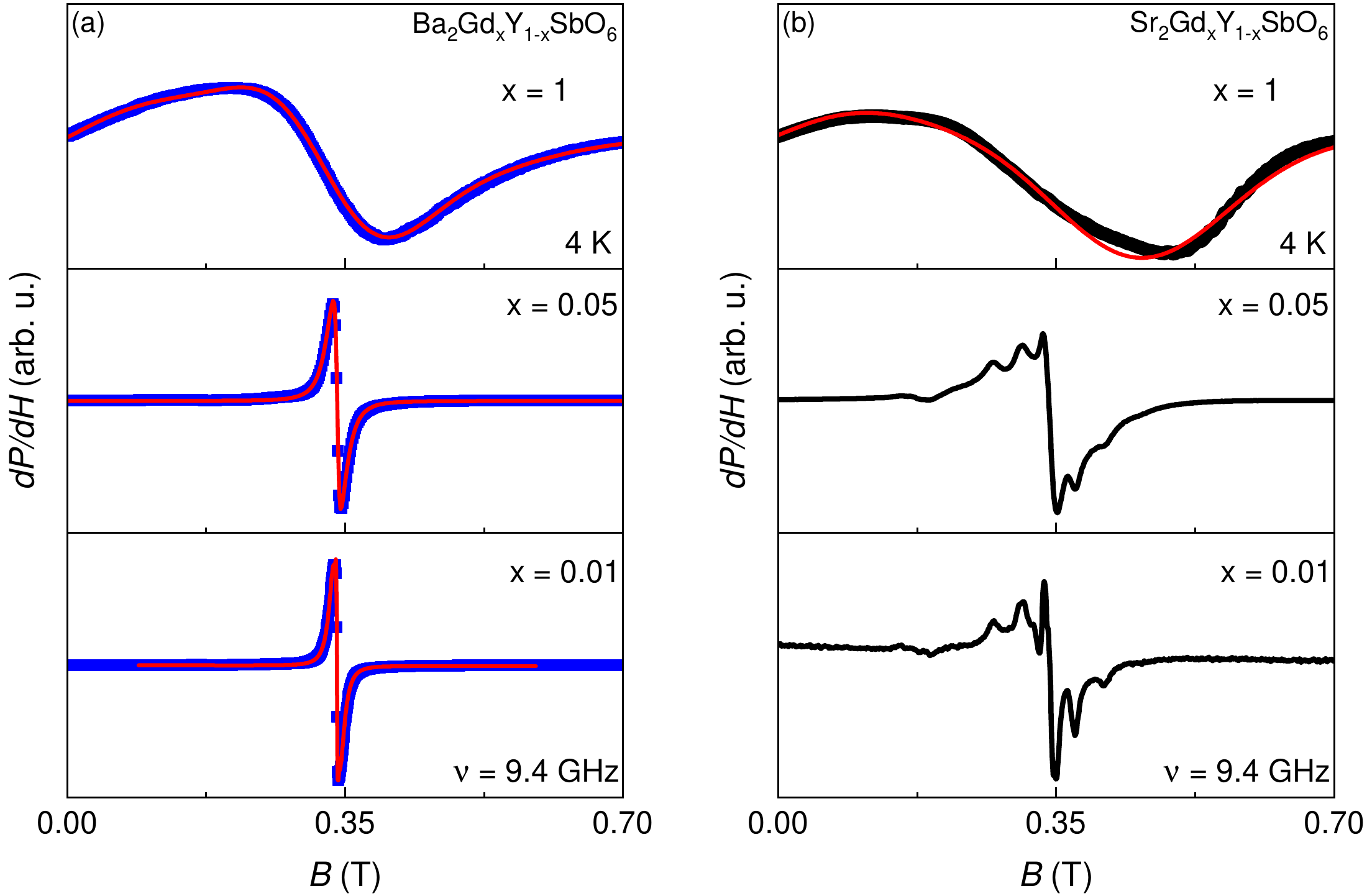}
\caption{ESR spectra of pure and diluted Ba$_2$Gd$_\mathrm{x}$Y$_\mathrm{1-x}$SbO$_{6}$ and Sr$_2$Gd$_\mathrm{x}$Y$_\mathrm{1-x}$SbO$_{6}$ taken at the X-band frequency at 4\,K. \label{sup:fig:ESR}}
\end{figure} 
\\ The absorption spectra of both compounds, shown in Figure \ref{sup:fig:ESR}, exhibit broad, asymmetrically distorted lines, which are optimally described by the superposition of the field derivatives of two Gaussian lines. This model provides a better fit than alternatives such as two Lorentzian lines or a powder pattern composed of Lorentzian resonances. The dominance of inhomogeneous broadening in these spectra arises from the distributions of local dipolar fields generated by neighboring Gd$^{3+}$ spins acting on each individual Gd$^{3+}$ ion and weak zero-field splitting of the $S =7/2$ octet into four doublets. Notably, the isotropic exchange interaction is weak in these systems, as evidenced by the absence of exchange narrowing effects. \\
The resonance fields and linewidths derived from spectral fits display only a weak dependence on temperature. For the Ba compound, the resonance fields of the two Gaussian components (with standard deviations representing linewidths) converge to 0.29/0.34\,T (0.14/0.07\,T) at elevated temperatures. In the Sr compound, the corresponding values are 0.22/0.38\,T (0.14/0.09\,T). The requirement for two Gaussian components in the fit reflects the anisotropy in dipolar couplings and crystal field effects, which are averaged over all orientations in the polycrystalline powder pattern. \\
To quantify the inhomogeneous broadening arising from nearest-neighbor dipolar fields, we apply the theoretical framework of Van Vleck \cite{VanVleck1957} (Equation 12):
\begin{equation}
    \langle \Delta\omega^2\rangle_{\mathrm{Av}} = \frac{3}{5} \left(\frac{2 \pi}{h}\right)^2 S(S+1) \sum_i \tilde{B}_{i}^2,
\end{equation}
where S = 7/2 and $\tilde{B}_{i} = g^2 \mu_\mathrm{B}^2 / d_{i}^3$, where $g$ stands for the Landé~factor and $d_i$ is the distance between nearest neighbors (see Table \ref{sup:tab:exchange}). This yields estimated broadening values of 0.0916\,T (10.6\,$\mu$eV) for \bgso{} and 0.0976\,T (11.3\,$\mu$eV) for \sgso{}, which are consistent with the observed order of magnitude of the dipolar couplings (see Table \ref{sup:tab:exchange}). \\
To isolate the contribution of zero-field splitting due to the crystal electric field, we conducted comparative ESR measurements on magnetically diluted samples, where the majority of Gd$^{3+}$ ions were substituted with non-magnetic Y$^{3+}$, leaving only 5 \% or 1 \% Gd$^{3+}$. The resulting spectra are presented in the middle and lower panels of Figure \ref{sup:fig:ESR}. Due to the dilution the influence of the dipolar fields can be neglected. \\
For \bgso{}, the spectrum of the diluted sample shows a significantly narrower absorption compared to the pure material. The resonance is well described by the sum of two Lorentzian lines centered at $B$~=~0.34\,T, corresponding to a g-factor of 1.99, in excellent agreement with the spin-only value of $g$~=~2 expected for the half-filled 4f shell of Gd$^{3+}$. The two Lorentzian components account for the residual anisotropy averaged in the powder pattern. \\
In contrast, the Sr compound exhibits a distinct splitting into at least seven lines in the diluted sample, arising from zero-field splitting due to a local uniaxial distortion at the Gd$^{3+}$ site. Given that the powder pattern is dominated by field orientations perpendicular to the uniaxial distortion, the crystal field parameter b$_2^0$ can be estimated from the spacing between adjacent resonance lines \cite{Schlott1988}. With a spacing of $\sim$ 0.03\,T, we determine b$_2^0~\approx$~40\,mK, which accounts for the additional broadening observed in the pure Sr compound relative to the Ba compound.

\section{Density-Functional Theory Calculations} \label{sup:sec:DFT} \noindent
Density-functional theory (DFT) band structure calculations were performed in the \texttt{FPLO} code~\cite{fplo} for the crystal structures of Ba$_2$GdSbO$_6$ and Sr$_2$GdSbO$_6$ determined experimentally. The PBE functional was chosen~\cite{pbe96}, and the DFT+$U$ correction was applied to Gd $4f$ states (on-site Coulomb repulsion $U=10$\,eV, Hund's coupling $J_H=1$\,eV). Magnetic interactions parameters were extracted by a mapping procedure~\cite{xiang2011}. \\
The results shown in Table~\ref{sup:tab:exchange} reproduce the experimental trend of weaker couplings in monoclinic \sgso{} compared to cubic \bgso{} and of the AFM nature of exchange couplings in both compounds.
\begin{table}[h]
\caption{\label{sup:tab:exchange}
Nearest-neighbor Gd--Gd distances $d_i$ (in\,\r A) and exchange couplings $J_i$ (in\,mK) according to DFT calculations, as well as the average dipolar coupling D (in\,mK, normalized to $S$~=~7/2 of Gd$^{3+}$) in \bgso{} and \sgso{}}
\begin{tabular}{ccc@{\hspace{1cm}}ccc}
 \multicolumn{3}{c}{\bgso{}} & & \multicolumn{2}{c}{\sgso{}} \smallskip\\
 & $d_i$ & $J_i$ & & $d_i$ & $J_i$ \smallskip\\
 \hline
 $J_1$ & 5.993 & 21.4 & $J_{11}$ & 5.850 & 0.5 \\
                  & & & $J_{12}$ & 5.862 & 6.7 \\
                  & & & $J_{13}$ & 5.878 & 2.6 \\
                  & & & $J_{14}$ & 5.893 & 11.4 \smallskip\\
\hline 
& & $D$ & & & $D$  \\
\hline
& & 11.6 & & & 12.4 \\
\hline
\end{tabular}
\end{table}
\\Whereas $\bar J=5.1$\,mK in \sgso{} is in a good agreement with the experimental value of 6.3\,mK from the Curie-Weiss temperature, the calculated exchange coupling in \bgso{} appears to be overestimated (21.4\,mK in DFT vs. 9.5\,mK experimentally). One reason for this discrepancy could be a weak structural deformation that can potentially appear in \bgso{} at low temperatures and reduce the exchange coupling with respect to its value obtained for the room-temperature cubic structure. \\
Using the experimental value of $\bar J=9.5$\,mK in \bgso{}, we estimate saturation field of the fcc antiferromagnet, $B_s=16\bar JS\,(k_\mathrm{B}/g\mu_\mathrm{B})=0.40$\,T in a good agreement with the experimental phase boundary of \bgso{} that reaches 0.39\,T at 60\,mK and extrapolates to around 0.40\,T at zero temperature. A similar estimate for \sgso{} returns $B_s=0.26$\,T, whereas magnetic order disappears at a much lower field of 0.15\,T experimentally. The deformation of the fcc lattice should thus play a role. Indeed, in \sgso{} 12 nearest-neighbor couplings split into $2\times J_{11}+4\times J_{12}+4\times J_{13}+2\times J_{14}$. According to DFT, the two leading couplings are $J_{12}$ and $J_{14}$ that together form an anisotropic triangular lattice with helical magnetic order in the regime of $J_{12}<2J_{14}$. The saturation field in this case is given by
\begin{equation}
 B_s=S\left(4J_{12}+4J_{14}+\frac{J_{12}^2}{J_{14}}\right)\left(\frac{k_\mathrm{B}}{g\mu_B\mathrm{B}}\right)
\end{equation}
and equals 0.20\,T for the exchange couplings from Table~\ref{sup:tab:exchange}. This simplified estimate shows that deformation of the fcc lattice not only reduces $\bar J$ but also lowers the saturation field with respect to $\bar J$ because of the reduced connectivity. \\
The traceless form of the dipolar coupling tensor eliminates the contribution of $D$ to the Curie-Weiss temperature measured on a powder sample. Therefore, our experimental estimates of $\bar J$ directly measure the average exchange coupling and correspond to the coupling energy of $\bar JS^2=10.0\,\mu$eV for $\bar J=9.5$\,mK in \bgso{}. This exchange energy is comparable in magnitude to the dipolar coupling energy of 12.2\,$\mu$eV for the Gd--Gd distance of 5.99\,\r A.

\section{Quasi-Adiabatic Demagnetization Measurements} \label{sup:sec:ADR} 
\begin{figure}[htb]
\includegraphics[width=0.8\linewidth]{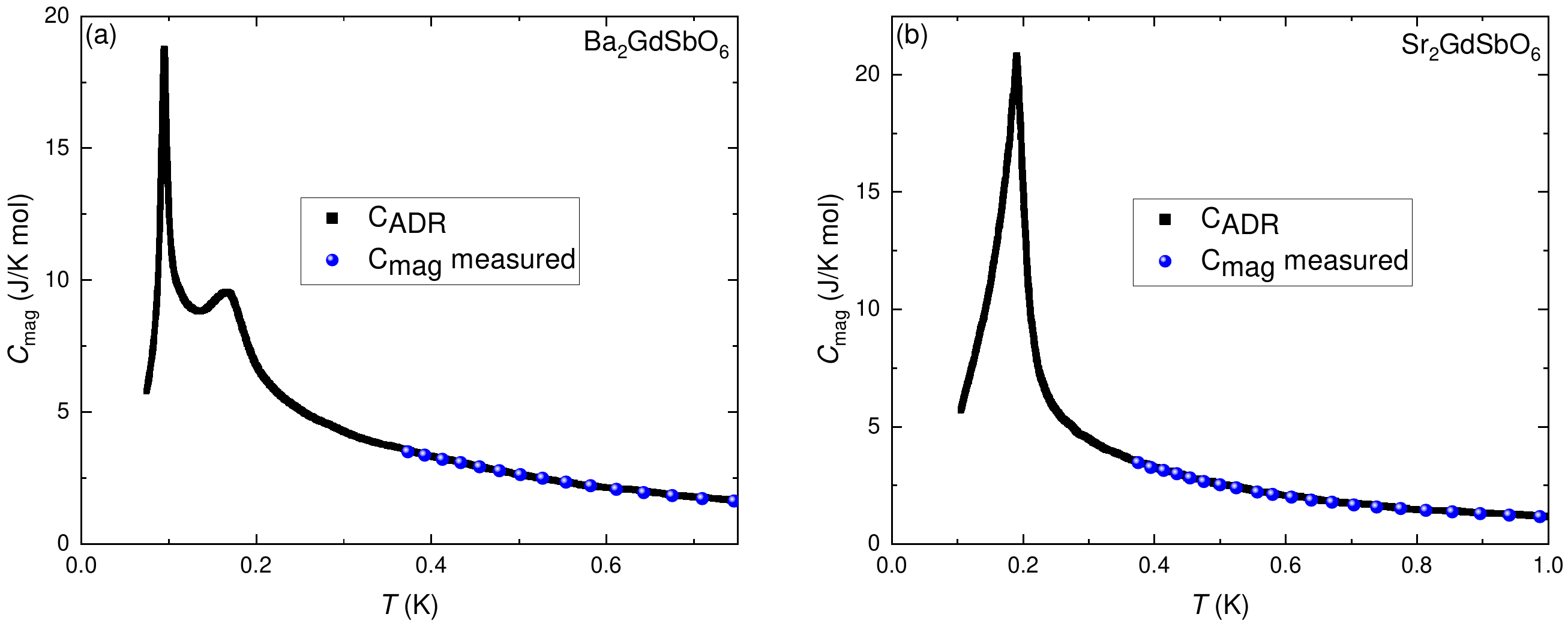}
\caption{Temperature-dependent heat capacity for (a) \bgso{} and (b) \sgso{} calculated from the ADR warming curve by Equation \ref{sup:eq:ADR} and the directly measured heat capacity in the PPMS. Good agreement was obtained for 110 and 100 nW respectively. \label{sup:fig:HC_ADR}}
\end{figure}

\noindent
In the ADR experiment, cylindrical pellets with a diameter of 15\,mm and an approximate weight of 3.3\,g and a thickness of roughly 3\,mm were utilized for each compound. These were formed by uniaxially pressing equal weights of sample material with silver powder (average grain size of 1\,µm). The pellets were installed into the same setup as \cite{Klinger2025}, which is an improved version previously used in \cite{Tokiwa2021, Jesche2023, Arjun2023, Arjun2023a, Telang2025, Treu2025}. The sample temperature $T(t)$ during cooldown and warming back to the starting temperature, was measured using a custom-built thermometer, which utilized a ruthenium oxide chip resistor. To enhance thermal coupling, the substrate thickness was reduced by approximately 80 \%. The thermometer was calibrated against a known reference thermometer, and resistance readings were taken with a Lake Shore Model 372 AC resistance bridge, integrated with a Model 3726 scanner, operating at a constant current of 1\,nA. The thermometer was affixed directly to the silver-sample pellets using GE varnish, with superconducting NbTi wires providing the electrical contacts. A (quasi-)adiabatic condition was achieved by evacuating the sample chamber using a home-made cryopump, to an ultra-high vacuum of less than 1.3\,mPa, after reaching the initial state at $T$~=~2\,K and $B$~=~5\,T. The applied magnetic field was ramped to zero and non-zero values at a rate of 5\,mT/s. To improve measurement accuracy, a second resistive chip, with a nearly temperature-independent resistance of 100\,$\Omega$, was affixed to the pellet to the opposite side of the pellet with respect to the thermometer using GE varnish. After the field ramped stopped, this chip was driven by a constant current, introducing a known electrical heat input. In a second set of experiment, the magnetic field was repeatedly sweeped between 0 and 1\,T after demagnetizing from the same initial conditions.
\begin{figure}[htb]
\includegraphics[width=0.8\linewidth]{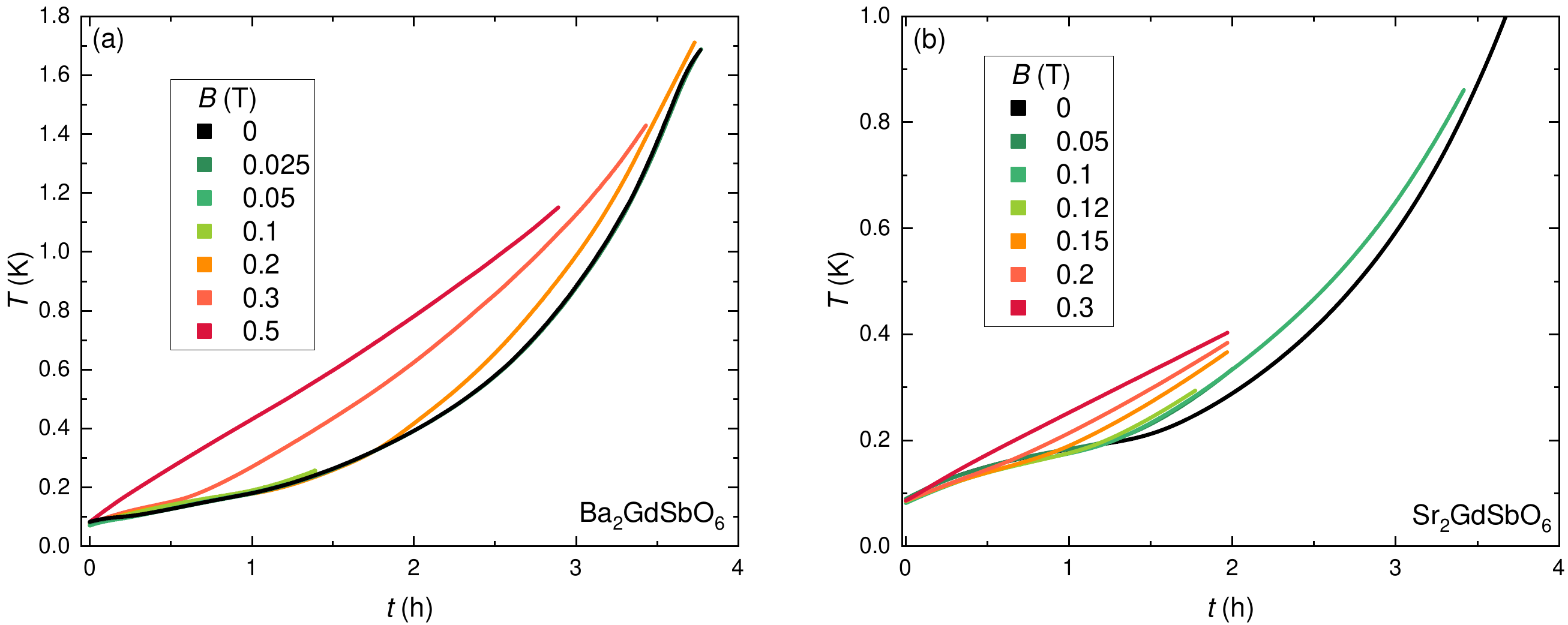}
\caption{Measured warmup curves for (a) \bgso{} and (b) \sgso{} after ramping the magnetic field from initial conditions at 2~K an $B$~=~5~T to zero and finite values at a rate of 5 mT/s, from which the heat capacity $C_{\mathrm{ADR}}$ was derived using Eq. \ref{sup:eq:ADR}. \label{sup:fig:ADR_warmup_curves}}
\end{figure}  \\
The heat capacity from the ADR warm-up curves was calculated using:
\begin{equation} \label{sup:eq:ADR}
C_{\mathrm{ADR}} = \frac{m_\mathrm{mol}}{ m }\frac{\mathrm{d} Q}{\mathrm{d} T} = \frac{m_\mathrm{mol}}{m} \frac{\dot{Q}}{\dot{T}},
\end{equation}
where $m$ is the mass of the magnetocaloric material, $m_\mathrm{mol}$ is its molar mass, $\dot{Q}$ is the heat input per unit time, and $\dot{T}$ is the time derivative of $T(t)$ during warm-up. $\dot{Q}$ consists of a known electrical contribution $Q_\mathrm{el} = RI^2$ ($\approx$\,1~\textmu W) from the second resistive chip and the residual parasitic heat leakage $Q_\mathrm{para}$ ($\approx$~100\,nW \cite{Klinger2025}) from residual gas, wiring, and radiation. The electrical contribution dominates $\dot{Q}$, minimizing the impact of the simplifying assumption that $Q_\mathrm{para}$ is temperature-independent, which also worked well \cite{Jesche2023, Telang2025, Treu2025}. The value of $Q_\mathrm{para}$ was determined by scaling the zero-field $C_\mathrm{ADR}$ curves to microcalorimetry results (see Figure~\ref{sup:fig:HC_ADR}), yielding 110\,nW for \bgso{} and 100\,nW for \sgso{}). These values were applied to both zero-field and in-field measurements, as no significant differences are expected between consecutive runs performed under identical conditions. Minor discrepancies between the compounds may arise from variations in sample mounting, thermometer and wire attachment and vacuum quality, as discussed previously \cite{Jesche2023, Telang2025}.
\\Additionally, approximately isentropic (d$S$ = 0) temperature traces $T(B)$ were measured with the heater turned off. Starting from $T$~=~2\,K and $B$~=~5\,T, the magnetic field was ramped to 1\,T, after which the ramp rate was reduced to 1.5\,mT/s, to increase the accuracy of data collection. Upon reaching zero field, the field was repeatedly ramped between 0\,T and 1\,T, with short waiting periods in between. During these intervals, the heater was briefly activated to increase the separation between the $T(B)$ traces. For each isentrope, the pair $(T, B)$ at the zero crossing of the magnetic Grüneisen parameters $\Gamma_\mathrm{mag}$ for each up- and down field ramp was determined where possible. We note in passing that the thermometer used in this measurement was not field-calibrated. While the effect is marginal at such low fields, an expected small negative magnetoresistance of below 1 \% below 0.5\,T for the RuO$_2$ thermometer would shift the temperatures upwards \cite{Watanabe2001}. Data acquisition was performed using the internal lock-in amplifier of the Lake Shore Model 372 AC resistance bridge, with data averaged over a 4-second time span (8\,s were used for the warmup curves). While this averaging has negligible effects on the warm-up curves due to their longer timescale (several hours), it can introduce noticeable changes in temperature and field for the $T(B)$ traces. This has been accounted for in the analysis of $\Gamma_\mathrm{mag}$.

\end{document}